\documentclass[12pt]{article}

\usepackage{amssymb}
\usepackage[centertags]{amsmath}
\usepackage{rotating}
\usepackage[dvips]{color}	

\usepackage{epsf}
\usepackage{pazh_eng}

\tightenlines

\newcommand{\me}{\mathop{\sf Me}\nolimits} 

\sloppypar

\begin{document}

{\renewcommand{\thefootnote}{\fnsymbol{footnote}}

\title{\bf Geometric Aspects and Testing of the Galactic Center Distance
Determination from Spiral Arm Segments
}

\author{\bf
I. I. Nikiforov\footnote{E-mail: \tt nii@astro.spbu.ru} and A. V. Veselova}

}
\setcounter{footnote}{0}

\vspace{-0.9em}
\begin{center}
{\it St.\ Petersburg State University, Universitetskii pr.\ 28,
\\

 Staryi Peterhof, St.\ Petersburg, 198504 Russia}
\end{center}

\vspace{-0.3em}

\sloppypar 
\vspace{2mm}
\noindent
{\bf Abstract}---We consider the problem of determining the geometric parameters of a Galactic spiral arm from
its segment by including the distance to the spiral pole, i.e., the distance to the Galactic center ($R_0$). The
question about the number of points belonging to one turn of a logarithmic spiral and defining this spiral
as a geometric figure has been investigated numerically and analytically by assuming the direction to the
spiral pole (to the Galactic center) to be known. Based on the results obtained, in an effort to test the
new approach, we have constructed a simplified method of solving the problem that consists in finding
the median of the values for each parameter from all possible triplets of objects in the spiral arm segment
satisfying the condition for the angular distance between objects. Applying the method to the data on the
spatial distribution of masers in the Perseus and Scutum arms (the catalogue by Reid et al. (2014)) has
led to an estimate of $R_0 = 8.8 \pm 0.5$~kpc. The parameters of five spiral arm segments have been determined
from masers of the same catalogue. We have confirmed the difference between the spiral arms in pitch
angle. The pitch angles of the arms revealed by masers are shown to generally correlate with $R_0$ in the
sense that an increase in $R_0$ leads to a growth in the absolute values of the pitch angles.


\clearpage

\section{INTRODUCTION}
\label{introduction}
Studies of the observed manifestations of the spiral
structure of galaxies, which give initial constraints
for constructing the theories of its formation, are of
fundamental importance for galactic astronomy and
stellar dynamics. Such studies for our Galaxy can
potentially give the most detailed information. However, they are complicated by the observation of the
spiral structure "edge-on" due to the location of the
Sun near the disk plane, by the fact that only the
local Galactocentric sector is well studied, by the
extinction in the disk, by the shielding of the region
behind the Galactic center, by the low accuracy or
the absence of data on the heliocentric distances for
spiral pattern tracers, and by other difficulties. There
exist two classes of methods for studying the Galactic
spiral structure: (1) based on the spatial distribution
of tracers of various features of this structure and
(2) based on the kinematics of such tracers (for more
details see, e.g., Nikiforov and Shekhovtsova 2001).
Each of the classes has its advantages and disadvantages. The kinematic approach in the case of
tying to the dynamical perturbations from the spiral pattern can produce well-conditioned parametric
models, but it requires additional assumptions (in
particular, about the nature of the spiral arms) that
may turn out to be inadequate. The spatial approach
has no need for such assumptions and is applicable
to any features of the spiral structure irrespective of
their nature, but reliable results from an analysis of
the spatial distribution may be expected only for those
types of tracers by which these features are identified
unambiguously.

Until recently, any spatial modeling of the Galactic
spiral structure has been based predominantly on the
data on tangential and other concentrations of gas
and objects tracing the spiral arms in an attempt to
"assemble" a regular structure from these individual
segments while guided by not so much by the geometry of individual concentrations as by their positions
(Efremov 2011). Of course, this can be done only
under some assumptions. Not always, but usually,
the following was assumed in such modeling: (1) the
arms have the shape of a logarithmic spiral; (2) the
spiral pitch angle ($i$) is constant; (3) the angle $i$
is the same for all arms; (4) the number of arms
in the Galaxy is specified (as a rule, two or four,
both variants were occasionally considered), which
largely defines the pitch angle; (5) the pole of the
spiral arms lies at the Galactic center; and (6) the
solar Galactocentric distance ($R_0$) is taken to be fixed.
Heterogeneous, including not very reliable, estimates
of the heliocentric distances with a large fraction of
kinematic distances, especially for distant structural
features, were often used in this case. The problem
is also complicated by the fact that the objects that,
in principle, gravitate toward the spiral arms also
populate the interarm space.

Obviously, due to these difficulties and, possibly,
the inadequacy of the assumptions made, it is impossible to obtain unambiguous results within this
approach. For example, Efremov (2011) concluded
that in the inner Galaxy ($R < 9- 10$~kpc) a four-armed spiral pattern with $i = 10^\circ -12^\circ$ is in better
agreement with the data on the distribution of neutral,
molecular, and ionized hydrogen. By contrast, Francis and Anderson (2012) conclude that the distribution of H\,I,
 giant molecular clouds, H\,II regions, and
2MASS sources corresponds to a two-armed logarithmic spiral with $i \sim 5\fdg5$ (also in the inner Galaxy,
$R < 12-15$~kpc). Pohl et~al. (2008) also preferred a
two-armed model, but with $i \sim 12^\circ$ , to represent the
CO data, while abandoning, however, the reproduction of some features in the constructed molecular gas
distribution by the model.

At the same time, an alternative approach is
possible --- the spatial modeling of a \emph{separate}\/ arm
segment or several arm segments \emph{separately}. In
particular, this makes it possible to determine the
segment pitch angles without the assumption (number~4 in our list) about the number of arms in
the Galaxy. Such an analysis was undertaken by
Pavlovskaya and Suchkov (1984), Avedisova (1985),
Dame et al.\ (1986), Vall{\'e}e (1988), and Grabelsky
et al.\ (1988), but the development of this approach
was restrained by a~shortage of reliable distance
estimates, especially for tracers with a strong concentration to the spiral arms.

The situation is changed noticeably by the recent
appearance of databases that allow one to produce
samples of spiral pattern tracers that (1) are strongly
concentrated to the spiral arms, (2) have at least internally accurate heliocentric distances, and (3) allow
the arm segments to be traced over a length long
enough to reveal their geometry (preferably the curvature). Samples with such properties are available
at least for masers with trigonometric parallaxes (see,
e.g., Reid et~al. 2014) and young open clusters (see,
e.g., Popova and Loktin 2005; Nikiforov and Kazakevich 2009).

An analysis of the spatial distribution of tracers
based on new data aimed at determining the geometric parameters of individual spiral segments has
 already been performed by a number of research groups.
Popova and Loktin (2005) estimated the pitch angle
of spiral arms based on data from the Homogeneous
Catalogue of OSC parameters without using assumption~4 but assuming the pitch angles to be identical for all segments (assumption 3 in our list). Reid et~al. (2009, 2014), Xu et~al. (2013), and Bobylev and
Bajkova (2013, 2014) performed the corresponding
modeling for masers with trigonometric parallaxes by
abandoning both assumptions~3 and~4. Note that
when investigating the nature of the Local arm (Xu
et al.\ 2013), which implies finding the parameters of
precisely this feature, the statement of the problem
itself requires abandoning assumption~3.

A similar analysis for classical Cepheids is more
problematic due to their generally less pronounced
concentration to the spiral arms. Popova (2006)
and Dambis et~al.\ (2015) applied the procedures
for identifying the objects belonging to the rigdelines of spiral segments to obtain more stable results
from Cepheids (note that Popova and Loktin (2005)
also used the same procedure for open clusters).
On the other hand, the more disperse distribution
of Cepheids is partly compensated for by the high
internal accuracy of their distance estimates and the
large size of their present-day samples. In both papers the results were obtained without assumption~4,
while in Dambis et~al.\ (2015) assumption~3 was also
abandoned.

If it is assumed that with the currently available
data the spatial modeling of spiral segments allows
their geometry to be established with confidence (in
particular, the pitch angle to be determined) without
relying on assumptions~3 and~4, then one can attempt
to go even further and to free the parameter $R_0$, i.e.,
to abandon assumption~6. This will make it possible
to model the spiral segments more completely, because from general considerations one might expect
the existence of a significant dependence of the pitch
angles on $R_0$ (as confirmed by our calculations). Furthermore, this can give a new method of determining
the distance to the Galactic center as the distance
from the Sun to the pole (geometric center) of the
spiral structure. The proposed method can become
the first spatial method of determining $R_0$ applicable to objects of the flat Galactic subsystem. The
method can be both absolute, when using the data on
masers with trigonometric parallaxes, and relative, if
it is applied to objects with photometric distances (for
the classification of $R_0$ measurements, see the review
by Nikiforov (2004)). In principle, the method can
be applied to any objects tracing the Galactic spiral
structure.

A rigorous consideration of this problem by taking
into account two uncertainties (the scatter across
the arm and the random errors in the heliocentric
distances) will possibly allow some other assumptions from the above list to be additionally abandoned.
However, the rigorous method suggests laborious calculations even in the variant being considered here 
(when abandoning only assumptions 3, 4, and~6). 
Therefore, first of all, we should check whether the 
  new approach is operable in principle both for the currently available data and in prospect. In this paper we 
  initially consider the idealized (geometric) problem of 
  reconstructing the parameters of a logarithmic spiral 
  as a figure from points belonging to it by assuming 
    that the direction to the spiral pole (Galactic center) 
    is known (Section~2). Then, to test the proposed 
    approach, we construct a simplified method of solving the problem for real data and apply it to masers 
      (Section~3). The simplified method is tested through 
        numerical simulations in Section~4. The method and 
        results obtained are discussed in Section~5. 

\section{THE PROBLEM OF DETERMINING THE 
PARAMETERS \\ OF A LOGARITHMIC SPIRAL 
FROM POINTS OF ITS SEGMENTS 
}
\subsection{The Formulas to Determine the Parameters 
of a Spiral from Three Points 
}

Consider the geometric problem on the possibility of reconstructing the parameters of a logarithmic 
spiral from points belonging to its segment. We will 
assume the spiral to be in the Galactic plane and 
the direction from the Sun to the spiral pole (to the 
Galactic center) to be known. The separate Galactic 
arm will then be represented by a segment of the 
logarithmic spiral 
\begin{equation}
R(\lambda) = |R_0|e^{k(\lambda - \lambda_0)}. \label{eq}
\end{equation}
Here, $\lambda \in  ( - \infty, + \infty )$ is the Galactocentric longitude (measured from the sunward direction clockwise
when viewed from the North Galactic Pole, i.e., in the
direction of Galactic rotation); $k \equiv \tan{i}$, where $i$ is
the pitch angle (negative for a trailing spiral); $\lambda_0$ is the
position parameter ($R(\lambda_0) = |R_0|$). In the sunward
direction $\lambda = 0 \pm 2\pi n, n \in \mathbb{Z}$.

Since the logarithmic spiral is specified by Eq.~\eqref{eq}
containing three parameters ($R_0, k, \lambda_0$), let us
consider the possibility of determining these parameters from three (different) points $M_1$ ($r_1, l_1, b_1$),
$M_2$ ($r_2, l_2, b_2$), and $M_3$ ($r_3, l_3, b_3$) lying on one spiral
turn when projected onto the Galactic plane. Here,
$r_j$, $l_j$, and $b_j$ are, respectively, the heliocentric distance, Galactic longitude and latitude of point $M_j$,
$j = 1, 2, 3$. Let us assume that $\lambda_1 < \lambda_2 < \lambda_3$; the
Galactocentric longitudes of such points then satisfy
the inequality
\begin{equation}
\lambda_3 - \lambda_1 < 2\pi. \label{delta_la}
\end{equation}

Formula~\eqref{eq} gives the expressions for the Galactoaxial distances $R_j$ of points $M_j$ via the corresponding Galactocentric longitudes:
\begin{equation}\label{R_i}
  R_j = |R_0|e^{k(\lambda_j - \lambda_0)}, \quad j=1,2,3.
\end{equation}
On the other hand, $R_j$ can be found from the coordinates $r_j$, $b_j$, and $l_j$ of points $M_j$ depending on the
spiral parameter $R_0$:
\begin{equation}\label{R_i_gal}
R_j = \sqrt{R_0^2 + r_j^2\cos^2{b_j} - 2R_0r_j\cos{l_j}\cos{b_j}}, \quad j=1,2,3.
\end{equation}

Passing to the Cartesian coordinates
\begin{equation}\label{XYZ}
X_j = r_j\cos{l_j}\cos{b_j}\,,\qquad
Y_j = r_j\sin{l_j}\cos{b_j}\,,\qquad
Z_j = r_j\sin{b_j}\,,
\end{equation}
we will write the following expressions for the Galactocentric longitudes:
\begin{equation} \label{XY_la}
X_j = R_0 - R_j\cos{\lambda_j}\,,\qquad
Y_j = R_j\sin{\lambda_j}\,.
\end{equation}

Formulas~\eqref{XYZ} and~\eqref{XY_la} show that at fixed $r_j$, $b_j$, and
 $l_j$ and given $R_0$ we can unambiguously determine
only the {\it nominal\/} Galactocentric longitudes $\Lambda_j$ for
points $M_j$ such that
\begin{gather}
  -\pi \leqslant \Lambda_1  \leqslant \Lambda_2 \leqslant \Lambda_3 < \pi ; \label{La_123}\\
\sin{\Lambda_j} = \frac{Y_j}{R_j},\quad \cos{\Lambda_j} = \frac{R_0 - X_j}{R_j}, \quad j = 1, 2, 3.
\label{X_i_La_i}
\end{gather}
The {\it rotational\/} longitudes $\lambda \in (-\infty, +\infty)$ can differ
from the nominal longitudes $\Lambda$ by an integer number
of complete rotations. The Galactocentric longitudes
enter into the final formulas to calculate the spiral parameters $R_0$ and $k$ as the differences of the longitudes
of two points belonging (by the initial assumption) to
one spiral turn, which allows the rotational longitudes
to be replaced by the nominal ones in the formulas.

Based on equalities~\eqref{R_i}, we will set up the following
system of equations for the unknown parameters $R_0$
and $k$:
\begin{equation}\label{R1/R2}  
R_1/R_2 = e^{k(\Lambda_1 - \Lambda_2)},\qquad 
R_2/R_3 = e^{k(\Lambda_2 - \Lambda_3)}. 
\end{equation}
Let us express $k$, for example, from the first equality
in~\eqref{R1/R2}, then  $k = \ln (R_1/R_2)/({\Lambda_1 -
\Lambda_2})$, and substitute it into the second one. Taking the logarithm of
the latter, we obtain the equation for $R_0$
\begin{equation}\label{basic_ln}
(\Lambda_3 - \Lambda_2)\ln{R_1} + (\Lambda_1 - \Lambda_3)\ln{R_2} + (\Lambda_2 - \Lambda_1)\ln{R_3} = 0.
\end{equation}
All quantities in~\eqref{basic_ln} are functions of only $R_0$ and the
coordinates of points $M_j$. $R_j$ and $\Lambda_j$ are defined by
Eqs.~\eqref{R_i_gal} and~\eqref{X_i_La_i}, respectively.\footnote{We can save on calculations using the equation $(\Lambda_3 - \Lambda_2)\ln{R_1^2}
+ (\Lambda_1 - \Lambda_3)\ln{R_2^2} + (\Lambda_2 - \Lambda_1)\ln{R_3^2} = 0$ equivalent to Eq.~\eqref{basic_ln}. It will then suffice to find $R_j^2$ without taking
the square root. $R_j$ does not need to be known to find $\Lambda_j$.
} Equation~\eqref{basic_ln} was
solved numerically.

Once $R_0$ has been found from Eq.~\eqref{basic_ln}, the other
two spiral parameters are determined by any pair from
the following expressions:
\begin{gather}
   k = \ln\left(R_j/R_m\right)\left/({\Lambda_j - \Lambda_m})\right.,\quad j, m = 1,2,3, \enskip j \ne m; \label{k_}\\
 \lambda_0 = \Lambda_j - \frac{\ln({R_j/|R_0|})}{k}, \quad j = 1, 2, 3 \label{l_0}.
\end{gather}
At $k = 0$ (circumference) Eq.~\eqref{basic_ln} is trivial, while
Eq.~\eqref{k_} is meaningless. Equation~\eqref{basic_ln} and expressions ~\eqref{k_}, \eqref{l_0} constitute the formal apparatus of the
method for determining the parameters of a logarithmic spiral that we will call the {\it three-point\/} one.

\subsection{The Number of Roots of the Three-Point
Equation for the Parameter $R_0$}

Since Eq.~\eqref{basic_ln} is transcendental, the question
about the number of its roots, i.e., the question about
the uniqueness of reconstructing the parameters of
a logarithmic spiral from three points belonging to
one its turn, deserves a separate consideration. Let
such points $M_1$, $M_2$, and $M_3$ belong to a spiral with
parameters $R_0$, $k$, and $\lambda_0$. Consider a~variable $R_0'$ that
has the meaning of a trial value of $R_0$ defining the
quantities $R_j$ and $\Lambda_j$, $j = 1, 2, 3$. For the left-hand
side of~\eqref{basic_ln} we will introduce the notation
\begin{equation}
 f(R_0') \equiv (\Lambda_3 - \Lambda_2)\ln{R_1} + (\Lambda_1 - \Lambda_3)\ln{R_2} + (\Lambda_2 -
 \Lambda_1)\ln{R_3}\,. \label{f_R_0'}
\end{equation}
At $R_0' = R_0$ the value of $f(R_0')$ is zero. But is the
reverse true: does finding such $R_0'$ that $f(R_0') = 0$ imply an \emph{unambiguous}\/ determination of the parameter
$R_0$ for the spiral passing through three points? I.e.,
do three points define a spiral in this statement of the
problem?

Our predominantly numerical study of the behavior of $f(R_0')$ depending on the initial data and spiral
parameters revealed basic properties of this function.
Let us illustrate them using the model spirals corresponding to the outer and inner (with respect to the
Sun) Galactic arms as an example.

Let the initial points $M_1$, $M_2$, and $M_3$ belong to
one turn of the logarithmic spiral representing the
Sagittarius arm with parameters $R_0 = 8.0$~kpc and
$i = -18\fdg7$ (Nikiforov and Shekhovtsova 2001), and
$\lambda_0 = -30^\circ$. In what follows, we will call this spiral the
{\it initial\/} one. When calculating the function $f(R_0')$ defined by the set of three initial points, we renumbered
the latter for each value of $R_0'$ in such a way that condition~\eqref{La_123} was fulfilled, i.e., the numbering of points
$M_j$ at $R_0' \ne R_0$ might coincide or not coincide with
the initial one. This ensures an increase of the point
number with growing formal longitude $\Lambda(R_0')$ at any
value of $R_0'$ for nondegenerate configurations (Λ$\Lambda_j \ne
\Lambda_m\,, j \ne m$). This numbering rule and condition~\eqref{La_123}
follow from the statement of the problem of searching
for a spiral passing through three specified points
in one turn and from the absence of an assumption
that the order in which the spiral passes the points is
known to us in advance

\begin{figure}[t!]
\centerline{%
\epsfxsize=15cm%
\epsffile{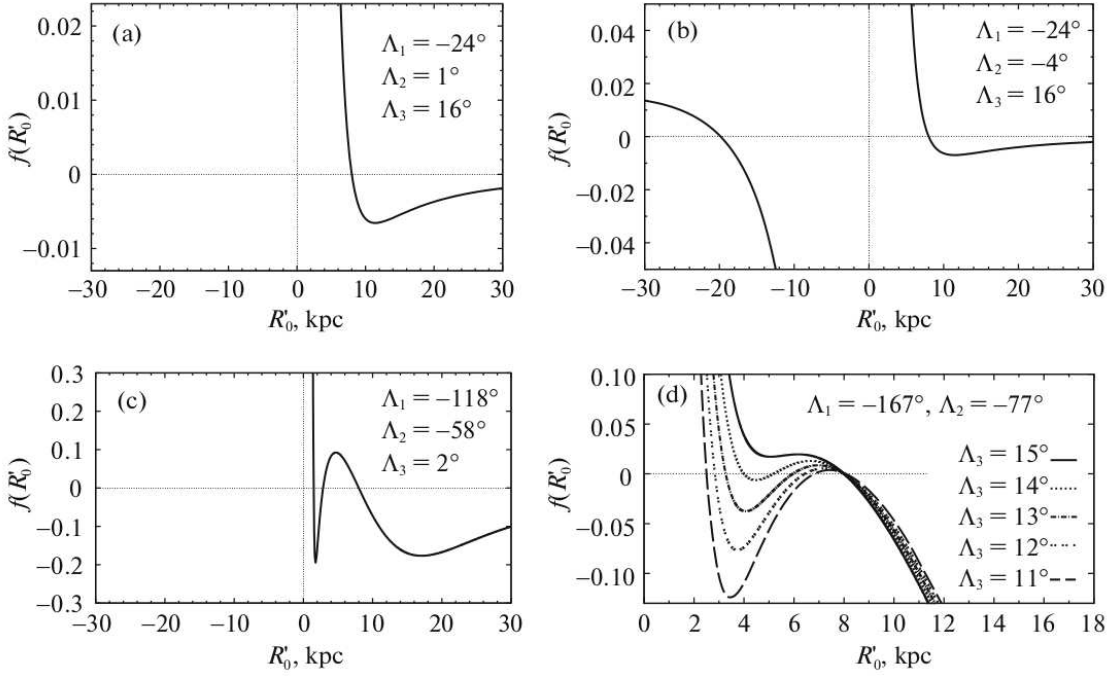}%
}
\caption{\rm Graphs of the functions $f(R_0')$ for segments crossing the direction to the spiral pole. (a)--(c) Examples of $f(R_0')$ with
a different number of roots. (d) An example of the dependence of the number of roots $f(R_0')$ on longitude $\Lambda_3$. The initial
numbering of points is indicated.}
\label{fR0_2_sides}
\end{figure}

\begin{figure}[t!]
\centerline{%
\epsfxsize=16cm%
\epsffile{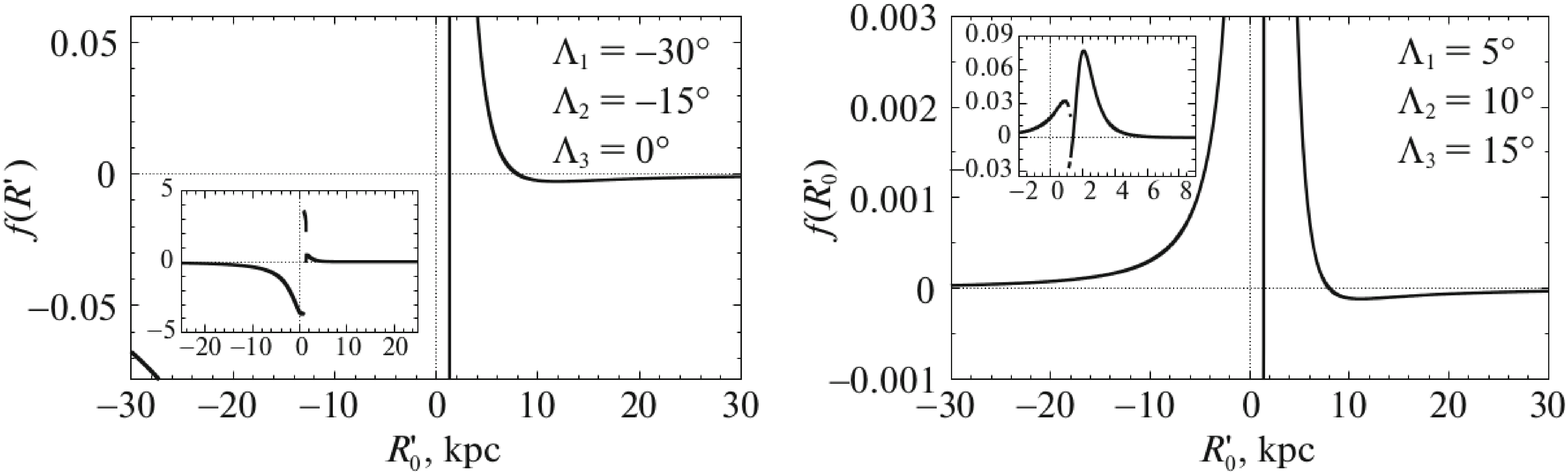}%
}
\caption{\rm Graphs of the functions $f(R_0')$ for segments located on one side of the direction to the spiral pole. The insets show
graphs of the same functions for larger intervals of values. In the neighborhoods of the roots $R_0' \approx 1.30$~kpc on the left panel
and $R_0' \approx 1.39$~kpc on the right panel $f(R_0')$ has a large, but finite, positive derivative. The initial numbering of points is
indicated.}
\label{fR0_1_side}
\end{figure}

When $R_0'\to \pm \infty$, the function  $f(R_0')\to 0$, as can
be seen from the following asymptotic expressions.
When $R_0'\to +\infty$ for any configuration of points and
when $R_0'\to -\infty$ for  $\Lambda_1 \leqslant 0$, $\Lambda_2<0$, and $\Lambda_3<0$,
\begin{equation}
    \label{fR0_as1}
    f(R_0') = [(Y_3-Y_2)X_1 +(Y_1-Y_3)X_2 +(Y_2-Y_1)X_3]/{R_0'^2} +o(1/R_0'^{2});
\end{equation}
when  $R_0'\to -\infty$,
\begin{gather}
    \text{ for }\Lambda_1 \leqslant 0, \Lambda_2>0,\Lambda_3>0\qquad
    f(R_0') = 2\pi (X_2-X_3)/{R_0'} +o(1/R_0');\label{fR0_as2}\\
    \text{ for }\Lambda_1 \leqslant 0, \Lambda_2<0,\Lambda_3>0\qquad
    f(R_0') = 2\pi (X_2-X_1)/{R_0'} +o(1/R_0').\label{fR0_as3}
\end{gather}
Here, the numbering of points (generally not the initial one) is set according to the above rule, the position of the spiral pole in the region of negative $X$ corresponds to a negative $R_0$. The sign of the functions
equivalent to $f(R_0')$ in all cases~\eqref{fR0_as1}--\eqref{fR0_as3} depends
only on the coordinates of points $M_j$, but not on $R_0'$.
This means that when $R_0'\to \pm \infty$, the function $f(R_0')$
asymptotically approaches the horizontal axis without crossing it. Consequently, the roots of Eq.~\eqref{basic_ln}
exist only in a limited interval of $R_0'$ (including the
initial $R_0$). This is confirmed by direct calculations
of $f(R_0')$ (see the examples in~\ref{fR0_2_sides} and~\ref{fR0_1_side}). The
function $f(R_0')$ at an argument close to $X_1$ or $X_2$ or
$X_3$ can undergo sharp oscillations and suffers discontinuities due to the change of the order in which the
points are numbered at configuration degeneracies
($\Lambda_j
= \Lambda_m\,, j \ne m$) (see the insets in Fig.~\ref{fR0_1_side}).

Our calculations show that $f(R_0') = 0$ is generally
reached at one, two, or three points. One root is fixed
($R_0' = R_0 = 8.0$~kpc), the {\it additional\/} roots change
their positions depending on the Galactocentric longitudes  $\Lambda_1$, $\Lambda_2$, and $\Lambda_3$\,.

If the spiral segment bounded by points $M_1$
and $M_3$ {\it crosses the $X$ axis\/} (the Galactic center--anticenter line), then the number of roots of the
equation $f(R_0') = 0$ depends on the arrangement of
points on the initial spiral. Figures~\ref{fR0_2_sides}a--\ref{fR0_2_sides}c  present
examples of the functions $f(R_0')$ for those configurations of points at which the number of roots is one,
two, or three. In special cases, the number of roots
is sensitive even to slight changes in the longitude
of one point, for example, to changes in $\Lambda_3$ if this
is the only positive longitude, $\Lambda_3$ is small, and the
interval between $\Lambda_j$ is large (Fig.~\ref{fR0_2_sides}d); note that for the
family $f(R_0')$ in Fig.~\ref{fR0_2_sides}d in the interval~$\Lambda_3 \in (14^\circ, 15^\circ)$
there exists a value at which the number of roots
is two.


In the case where the segment is located {\it on one
side of the $X$ axis}, including the cases where one
of its (extreme) points $M_1$ and $M_3$ is exactly on the
$X$ axis, there always exist two roots one of which
coincides with the initial $R_0$ (Fig.~\ref{fR0_1_side}).

\begin{figure}[t!]
\centerline{%
\epsfxsize=16cm%
\epsffile{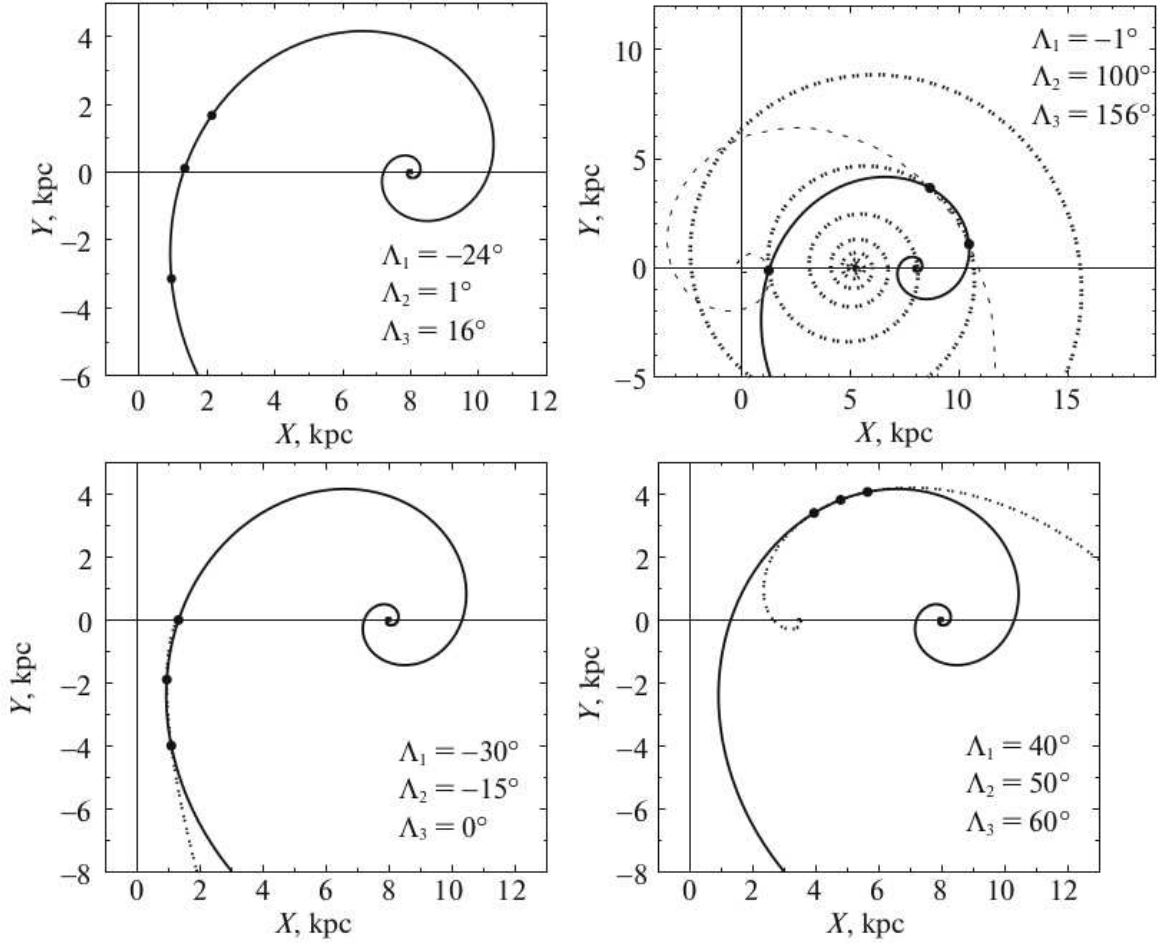}%
}
\caption{\rm Examples of the configurations of initial points at which the segment formed by them crosses the direction to the spiral
pole (upper panels) or is located on one side of this direction (lower panels). The solid line indicates the initial spiral. On the
lower left panel the pole of the additional spiral is near point $M_3$. The numbering of points is initial.}
\label{int_2+1_sides}
\end{figure}

A spiral passing through all three initial points
corresponds to each root of the function $f(R_0')$ and
the two remaining parameters calculated for it from
Eqs.~\eqref{k_} and~\eqref{l_0} (see the examples in Fig.~\ref{int_2+1_sides}), i.e.,
the number of such spirals is equal to the number of
roots. Thus, generally, the logarithmic spiral passing
through three points in one turn cannot be unambiguously determined from them even if the direction
to the spiral pole is known.

We will consider the revealed trends in more detail
using the case (important for some applications) of
initial points $M_j$ equidistant in longitude $\Lambda$ as an
example. Denote the parameter $R_0$ of the initial spiral
by $R_{0,1}$ and the additional roots of~\eqref{basic_ln}, if they exist,
by $R_{0,2}$ and $R_{0,3}$. Let us investigate the dependence
of the presence/absence of additional roots and quantities $R_{0,2}$ and $R_{0,3}$ on the positions of the triplets
of initial points $\big\lbrace M_{1,n}, M_{2,n}, M_{3,n} \big\rbrace_{n=1}^{N}$\, lying on the
initial spiral. We will specify the initial longitudes of
these points as follows:
\begin{eqnarray}
\label{scan}
\begin{split}
 &\Lambda_{2,n} - \Lambda_{1,n} \equiv \Delta\Lambda \equiv \Lambda_{3,n} - \Lambda_{2,n}\,, \\
 &\Lambda_{1,n+1} - \Lambda_{1,n} = 1^\circ,\\
 &\Lambda_{1,1} = -180\deg,\enskip \Lambda_{3,N} = 180\deg-1\deg=179\deg.
 \end{split}
 \end{eqnarray}
 Thus, the (almost complete) spiral turn was covered
by the triplets of points with a $1^\circ$ step. A set of
solutions was determined for each triplet.

\begin{figure}[t!]
\centerline{%
\epsfxsize=11cm%
\epsffile{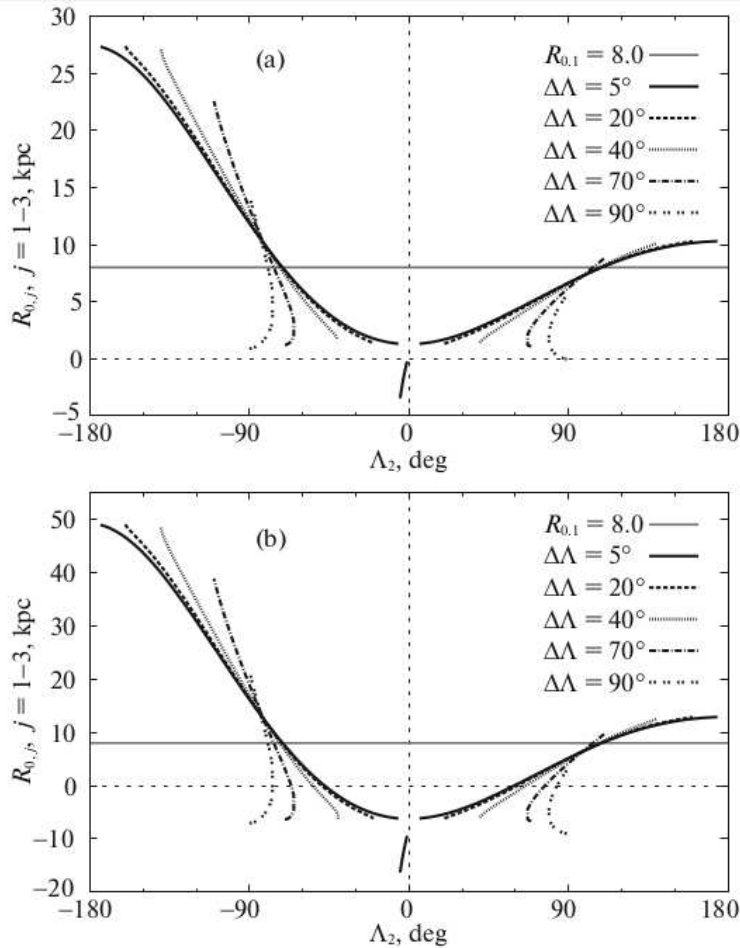}%
}
\caption{\rm Branches of solutions for the equation $f(R_0') = 0$ as functions of the longitude of the central point in the triplet at
different $\Delta\Lambda$ for the spirals representing the Sagittarius (a) and Perseus (b) arms.}
\label{R_d_la}
\end{figure}

\begin{figure}[t!]
\centerline{%
\epsfxsize=16cm%
\epsffile{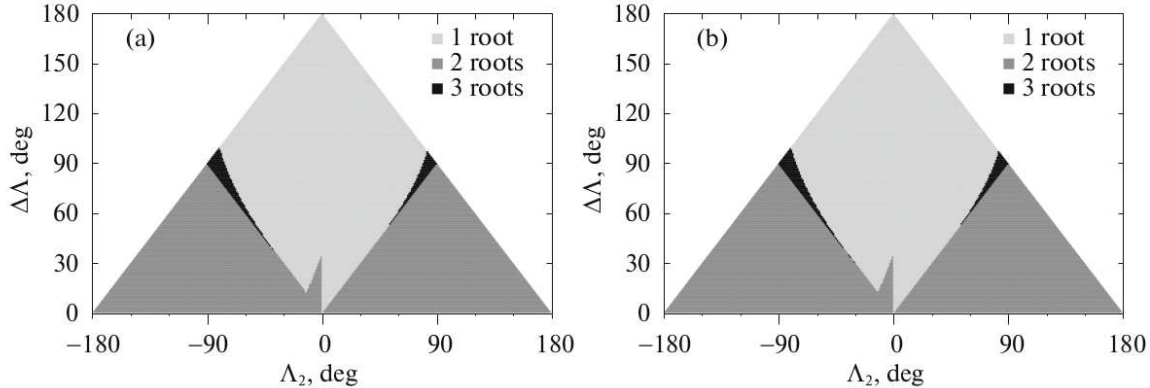}%
}
\caption{\rm Dependence of the number of roots of Eq.~\eqref{basic_ln} on the longitude of the central point $\Lambda_2$ and the distance between the adjacent points $\Delta\Lambda$ for the spirals representing the Sagittarius (a) and Perseus (b) arms.}
\label{false_Sgr_Per}
\end{figure}

In Fig.~\ref{R_d_la}a the roots of $f(R_0')$ are plotted against
the longitude of point $M_2$ at some values of $\Delta\Lambda$
for the model representing the Sagittarius arm.
The number of roots for the triplet of points with
specified $\Delta\Lambda$ and $\Lambda_2$ is equal to the number of
intersections of the straight line $\Lambda_2 = {\text {const}}$ with the
branches of solutions shown in the figure (including
the branch $R_{0,1} = 8$~kpc) for this $\Delta\Lambda$. A generalization of the results obtained is presented in Fig.~\ref{false_Sgr_Per}a for
the ($\Lambda_2$, $\Delta\Lambda$) configuration plane on which the region
of possible configurations (RPC) in our statement
of the problem is bounded by the isosceles triangle
with vertices ($\Lambda_2$, $\Delta\Lambda$) = $(-\pi, 0), (\pi, 0), (0, \pi)$. The RPC does not form a closed set, because the right
side of the triangle (by definition, $\Lambda_3 < \pi$) and its
base (by definition, $\Delta\Lambda > 0$) do not belong to it.
Analysis of the graphs for the branches of solutions
(Fig.~\ref{R_d_la}a) and the results of scanning according to the
rule~\eqref{scan} show that the RPC is divided into six two-dimensional regions all points of each of which have
the same number of roots (Fig.~\ref{false_Sgr_Per}a).

The region of a unique root ($R_{0,1} = 8$~kpc) occupying slightly less than half of the RPC area is
largest. The overwhelming majority of triplets whose
segments cross the $X$ axis, with all triplets with $\Lambda_2  =
0$ being among them, belong to it. The properties of
the region of a unique root turn out to be the same
for the models of other Galactic arms as well, i.e.,
the three-point configurations with a unique solution
of Eq.~\eqref{basic_ln} correspond to the arm segments in the
solar sector of the Galaxy. Such segments are usually
revealed by tracers with non-kinematic (i.e., more reliable) distances (see, e.g., Popova and Loktin, 2005;
Nikiforov and Kazakevich 2009; Dambis et al.\ 2015).

Two roots exist in the two large regions that are
bounded by the triangles with vertices $(\Lambda_2,\Delta\Lambda)=(-\pi,0), (0,0),
\left(-\frac{\pi}{2},\frac{\pi}{2}\right)$ and $(0,0), (0,\pi), \left(\frac{\pi}{2},\frac{\pi}{2}\right)$
and that occupy together exactly $\frac12$ of the RPC area
(Fig.~\ref{false_Sgr_Per}a). However, the segments lying on one side
of the $X$ axis, including the cases where the extreme point ($\Lambda_1 = 0$ or $\Lambda_3 = 0$) touches this axis,
correspond to them, and such configurations are not
typical of tracers with non-kinematic distances.

Between these regions of two roots and the region of a unique root there are small (in area) two-dimensional regions of three roots bounded, respectively, by the line $\Lambda_2=-\Delta\Lambda$ and the line $\Lambda_2=+\Delta\Lambda$
(Fig.~\ref{false_Sgr_Per}a). The first region is revealed at $ 39^\circ\leqslant\Delta\Lambda \leqslant 99^\circ$ for triplets with $\Lambda_{1,n} < -\Delta\Lambda < \Lambda_{2,n} < 0 < \Lambda_{3,n}$.
The second region is revealed at $54^\circ\leqslant \Delta\Lambda \leqslant 97^\circ$  for
triplets with $\Lambda_{1,n} < 0<\Lambda_{2,n} < \Delta\Lambda <
\Lambda_{3,n}$\,.

A small region of two roots that has a triangular
shape and is bounded by the lines  $\Lambda_2=0$ and $\Lambda_2=-\Delta\Lambda$, but does not include them, is also revealed at
 $\Delta\Lambda\leqslant 34\deg$ (Fig.~\ref{false_Sgr_Per}a). The region corresponds to short
segments crossing the $X$ axis with Λ$\Lambda_{1,n} <
-\Delta\Lambda < \Lambda_{2,n} < 0 < \Lambda_{3,n}$\,. However, even when $\Lambda_2\to-0$ for
fixed $\Delta\Lambda$, the additional root $R_{0,2}$ is much smaller
than the initial $R_0=8$~kpc, while it becomes even
smaller as $\Lambda_2$ decreases; the dependence $R_{0,2}(\Lambda_2$)
suffers a discontinuity at $\Lambda_2=-\Delta\Lambda$, and the solution~$R_{0,2}$ abruptly passes to another branch with
larger values (see the graph for  $\Delta\Lambda=5\deg$ in Fig.~\ref{R_d_la}a).
As $\Delta\Lambda$ increases, the branch $R_{0,2}(\Lambda_2$) at $-\Delta\Lambda<\Lambda_2<0$ goes abruptly into the negative region. Therefore, this branch fell into the interval of~$R_{0,j}$ shown in
Fig.~\ref{R_d_la}a only for~$\Delta\Lambda=5\deg$. As the lower limit of the
search for roots decreases, the height of the region
of two roots under consideration increases (but very
slowly) due to the upward shift of its upper inclined
boundary (in Fig.~\ref{false_Sgr_Per}a this region is given for $R'_{0}\geqslant-5000$~kpc). Obviously, the presence of such a branch
of solutions deviating strongly from the initial value
of~$R_{0,1}$\, does not produce any ambiguity in choosing
the root.

The same is also true for most of the length of
other branches of additional roots (Fig.~\ref{R_d_la}a), where the
solutions are distinguishable from the initial spiral in
pitch angle as well. The cases where the middle point
of the segment is near the traverse directions, $\Lambda_{2,n}$ from $-80^\circ$ to $-70^\circ$ and from $+100^\circ$ to $+115^\circ$ (the
intervals of intersections of the line $R_{0,1}=8$~kpc by
the branches of additional roots in Fig.~\ref{R_d_la}a), constitute
an exception. Such situations are quite possible when
analyzing the spiral arm segments using masers the
data on which cover predominantly quadrants~I and
II (e.g., Reid et~al. 2014; see also Section~3 in this
paper). This means that some algorithm for choosing
between the roots~$R_{0,j}$ is generally needed.

Finally, there exist two degenerate one-dimensional
regions (i.e., lines) of two roots to which the merging/bifurcation points of the branches of two additional roots, where these branches have infinite
derivatives, correspond in Fig.~\ref{R_d_la}a. In Fig.~\ref{false_Sgr_Per}a these
lines form the boundaries between the region of one
root and the regions of three roots. Having zero
area, the lines of two roots for an arbitrary scanning
of type~\eqref{scan} are not revealed and, therefore, are not
displayed in Fig.~\ref{false_Sgr_Per}a.

The same study was performed for the model spiral
representing the Perseus arm with parameters $R_0 = 8.0$~kpc, $i = -18^\circ$, and $\lambda_0 = 97\deg$ (Nikiforov and
Shekhovtsova 2001). The results obtained (Figs.~\ref{R_d_la}b
and~\ref{false_Sgr_Per}b) show that the basic properties of the solutions
of Eq.~\eqref{basic_ln} remain as before when passing from the
inner spiral arm to the outer one. The differences
concern only some details. For the Perseus arm the
lower parts of all branches of additional roots lie in
the region $R_0 < 0$~kpc (because many of the triplets
near the outer arm have negative coordinates $X_j$),
the scatter of values of $R_{0,2}$ and $R_{0,3}$ turns out to
be larger, and the left region of three rots is revealed
at $\Delta\Lambda \geqslant 31^\circ$ (for the Sagittarius arm at $\Delta\Lambda \geqslant 39^\circ$).
The remaining parameters of the small regions of
two and three roots are the same as those for the
Sagittarius arm to within~$1\deg$.

On the whole, the changes of the results are insignificant when varying the pitch angles of the initial
spirals as well. The quantity $i$ affects only some characteristics of the small regions: the region of two roots
adjacent to the line $\Lambda_2=0$ is reduced with decreasing
$|i|$ (at $i=-10\deg$ for both arms $(\Delta\Lambda)_{\text{max}}=17\deg$); the lower (in $\Delta\Lambda$) boundary of the left region of three roots
slightly varies.

\subsection{On the Possibility of Determining the
Parameters \\ of a Logarithmic Spiral from Four Points
}
Since the three-point method generally yields an
ambiguous result, let us consider the possibility of
reconstructing the parameters of a spiral from four
points~$M_j$, $j=1,2,3,4$, located on one its turn. From
the quadruplet of points we will single out two arbitrary triplets  $M_1,M_2,M_3$ and $M_1,M_3,M_4$, write
equations of the form~\eqref{basic_ln} for them, and pose the
problem to find the same solution for both triplets.
This leads to the equation for $R_0$
\begin{equation}
\label{f4}
 f_4(R_0') \equiv |f_{123}| + |f_{134}|=0,
\end{equation}
where
\begin{gather}
f_{123} (R_0')= (\ln{R_1}-\ln{R_2})(\lambda_2 - \lambda_3) + (\ln{R_3}-\ln{R_2})(\lambda_1 - \lambda_2), \\
f_{134} (R_0')= (\ln{R_1}-\ln{R_3})(\lambda_3 - \lambda_4) + (\ln{R_4}-\ln{R_3})(\lambda_1 - \lambda_3).
\end{gather}
The moduli in~\eqref{f4} are needed to avoid the extraneous
roots at such $R_0$ that $f_{123}=-f_{134}\ne0$ to which the
spirals passing through four (and even three) points
do not correspond.

A check showed that Eq.~\eqref{f4} actually has a
unique root at the point where the positive-definite
function $f_4(R_0')$ {\it touches\/} the horizontal axis. However, even at a small variation in the position of at least
one point $M_j$ Eq.~\eqref{f4} generally becomes unsolvable,
because arbitrary four points do not lie on one spiral
turn. This makes the potential four-point method
inapplicable to pseudo-random or real data. This
property is retained as the number of initial points
increases further.

On the other hand, an arbitrary triplet of points
lies on one turn of at least one spiral, and, hence,
at least one solution will always exist when varying
the positions of the points. Therefore, the three-point
method can be applied to real data and in numerical
experiments provided that a criterion for choosing
between the roots is introduced. Below we will use
the three-point method.

\begin{table}[t!]
 \centering
 \tabcolsep0.75cm
 \caption{\label{tab_masers} \rm Data on the spiral segments identified by Reid
et al.\ (2014) based on masers. $l_{\text{min}}$, $l_{\text{max}}$, and $\Delta l$ are the
boundaries and the extent in Galactic longitude
}
 \vspace{5 pt}
 \begin{tabular}{l|c|c|c|c}
  \hline
  Segment & Number of masers & $l_{\text{min}}$ & $l_{\text{max}}$ & $ \Delta l$ \\
  \hline
  Scutum arm & 17 & $6^\circ$ & $32^\circ$ & $26^\circ$ \\
  Sagittarius arm  & 18 & $-9^\circ$ & $52^\circ$ & $61^\circ$ \\
  Local arm  & 25 & $60^\circ$ & $239^\circ$ & $179^\circ$ \\
	Perseus arm  & 24 & $43^\circ$ & $241^\circ$  & $198^\circ$\\
  Outer arm & 6 & $75^\circ$ & $196^\circ$ & $121^\circ$ \\[0.05cm]
    \hline
 \end{tabular} 
 \end{table}

\section{APPLYING THE THREE-POINT METHOD
OF DETERMINING \\ THE SPIRAL
PARAMETERS TO THE DATA ON MASERS
}

The formal apparatus to determine a spiral from
initial points is needed as a simple tool for studying
the possibilities of estimating $R_0$ from the geometry of
spiral arm segments as a function of problem parameters by the method of numerical simulations. Such a
study will be performed in the next paper.

However, the three-point method also allows the
new approach to be immediately tested on real data.
To solve this problem, we chose maser sources with
trigonometric parallaxes, because they undoubtedly
meet all three requirements for performing the spatial
modeling of individual arm segments formulated
in the Introduction and have absolute (needing no
calibration) heliocentric distance measurements. The
data were taken from the catalogue of H$_2$O and
CH$_3$OH masers in high-mass star-forming regions
(Reid et al.\ 2014). The catalogue contains the characteristics of $103$ masers from the BeSSeL, VERA,
VLBI, and EVN surveys; the mean parallax accuracy
is 20~$\mu$as. Based on these data, Reid et al.\ (2014)
identified five spiral segments, assigning the vast majority of masers to a particular segment by associating the positions of masers with CO and H\,I emission
features on the $l-V_{\text{LSR}}$ diagram, where $V_{\text{LSR}}$ is the
line-of-sight velocity relative to the local standard of
rest. In this paper we adopt the distribution of masers
over the segments proposed by Reid et al.\ (2014) (for
the observational characteristics of the segments, see
Table~\ref{tab_masers}).

\subsection{The Algorithm}

Each sample of masers assigned to one of the
segments was analyzed separately using an algorithm
based on the three-point method. The distance er-
rors for the maser sources were disregarded at this
stage (checking whether the approach is operable).
From the masers of each sample only those sets of
three objects $M_{1,j}$, $M_{2,j}$, and $M_{3,j}$ were selected
for which the angular distance between the adjacent
objects $l(M_{m+1,j}) -
l(M_{m,j})$, $m=1,2$ was no less
 than some adopted value $(\Delta l)_{\text {min}}$ (here, $j$ is the set
 number); $(\Delta l)_{\text {min}}$ was reoptimized for each sample.
 For each of the selected triplets of objects we calculated the values of $R_{0}$ as the roots of Eq.~\eqref{basic_ln} and
the parameters $k$ and $\lambda_0$ corresponding to them from
Eqs.~\eqref{k_}  and~\eqref{l_0}. $R_0$ was sought in the segment
$[-60.0,60.0]$~kpc. We considered two techniques for
 processing the cases of a nonunique solution for $R_0$:
(1) eliminating the triplets with a nonunique solution
and (2) choosing such a root from~$R_{0,1,j}$, $R_{0,2,j}$, and,
possibly, $R_{0,3,j}$ that leads to the smallest scatter of
sample masers relative to the spiral determined by
them. The first technique is simple and more reliable,
 but it is applicable only at a significant fraction of
  triplets with a unique solution in the total sample.
 Since the spiral arms turn out to be the trailing ones
from these data (Reid et al.\ 2014), the triplets corresponding to the leading spirals (with $k_j>0$) were
excluded from consideration in both techniques.

\begin{figure}[t!]
\centerline{%
\epsfxsize=16cm%
\epsffile{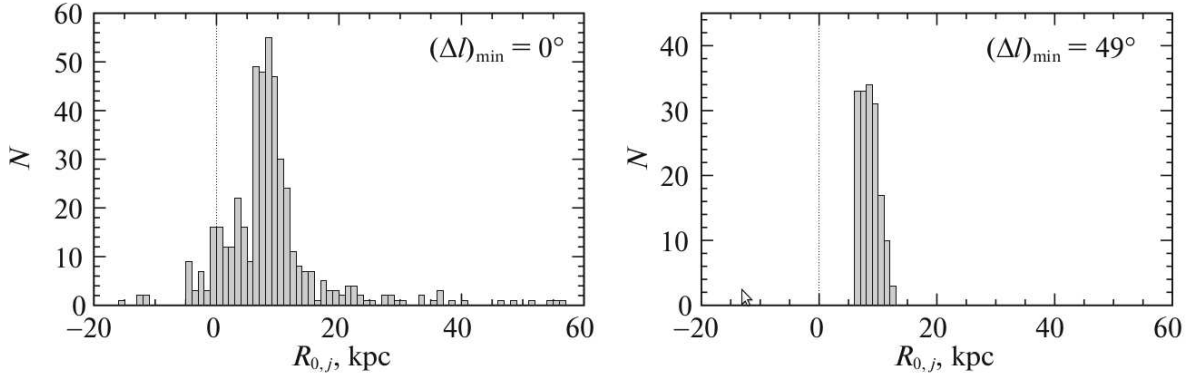}%
}
\caption{\rm $R_{0,j}$ distribution function for the Perseus arm at $(\Delta
l)_{\text {min}}=0\deg$ (the complete set of all possible triplets of masers) and at optimal $(\Delta l)_{\text{min}} = 49^\circ$.
}
\label{f_Per}
\end{figure}

At fixed $(\Delta l)_{\text {min}}$ the point estimates of the parameters $R_0$, $k$, and $\lambda_0$ for the segment under consideration
  were defined as the medians ($\me$) of the sets $\lbrace R_{0,j}\rbrace$,
 $\lbrace k_j\rbrace$, and $\lbrace\lambda_{0,j}\rbrace$ obtained from all of the triplets    $\lbrace
M_{1,j}$,~$M_{2,j}$,~$M_{3,j}\rbrace$ left after the eliminations by
  one of the above techniques; here, $j=1,\:\ldots,\:N_{\text{sol}}$\,,
 where $N_{\text{sol}}$ is the number of solutions involved in the
processing. As an optimal $(\Delta l)_{\text {min}}^0$ we then chose
$(\Delta l)_{\text {min}}$ for which the variance of the mean $\langle R_0\rangle$ was
smallest. As the final result for the segment sample
 we took the estimates of $\me R_0$, $\me k$, and $\me \lambda_0$
 found for $(\Delta l)_{\text {min}}^0$. We calculated the boundaries of
 min
the $1\sigma$ confidence intervals for these estimates based
  on order statistics (see, e.g., Kobzar' 2006).

For the {\it model spiral\/} (with the parameters $\me R_0$, $\me k$, and $\me
\lambda_0$ obtained at $(\Delta l)_{\text {min}}=(\Delta l)_{\text {min}}^0$\,) we determined the $1\sigma$ confidence region (see Appendix~A1). Below we give this region for the
Perseus arm; for other segments the boundaries of
the confidence regions, being less regular in pattern,
are not informative. For all segments we also made
an attempt to estimate the natural root-mean-square
(rms) scatter $(\sigma_{\text{w}})_0$ of masers across the spiral arm
(see Appendix~A2).

In this way we processed the samples of masers for
all five Galactic spiral arm segments identified by Reid
et al.\ (2014).

\subsection{Results for the Individual Spiral Segments
}
The most reliable estimates of the model parameters were obtained for the sample of 24 masers assigned to the Perseus arm. For this segment the
triplets of masers through which more then one spiral
passed were excluded from consideration. $(\Delta l)_{\text {min}}$ 
was varied from $0^\circ$ (the complete set of triplets was
used) to $70^\circ$.

Comparison of the $R_{0,j}$ distributions at various
values of $(\Delta l)_{\text
{min}}$ shows that as~$(\Delta l)_{\text {min}}$ increases,
the number of negative values of~$R_{0,j}$ decreases, as
does the variance of the distribution. For triplets with
$(\Delta l)_{\text {min}}
\geqslant 49^\circ$ the values of $R_{0,j}$ lie in the positive
region and are grouped in the segment $[6,13]~$kpc
(Fig.~\ref{f_Per}).

\begin{figure}[t!]
\centerline{%
\epsfxsize=14cm%
\epsffile{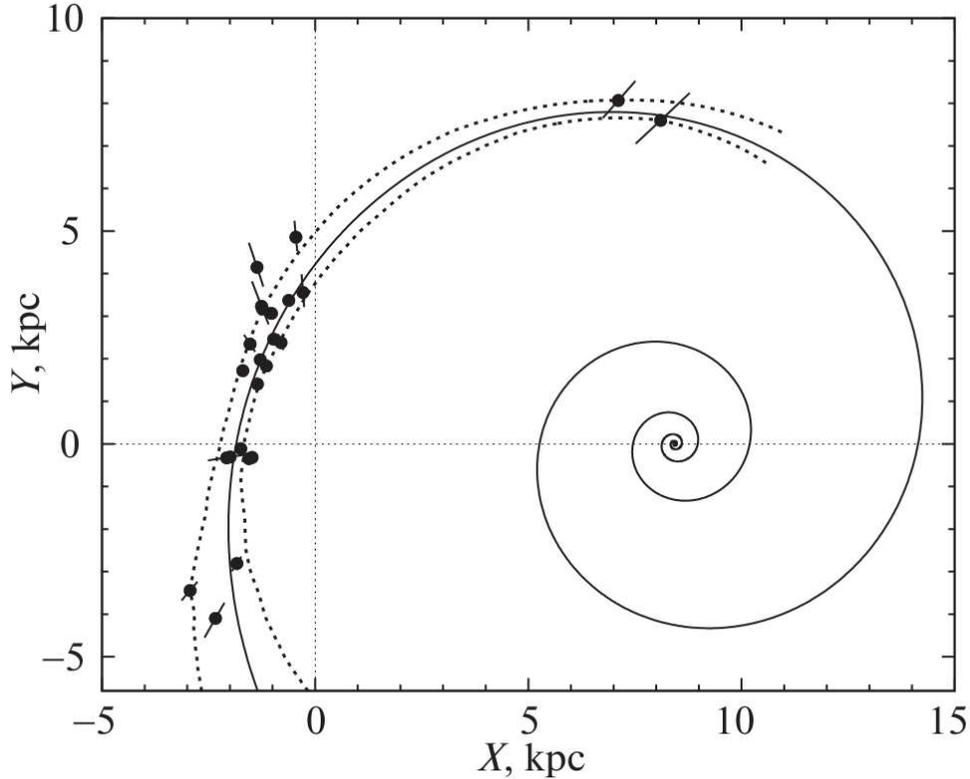}%
}
\vskip-0.3cm
\caption{\rm  The distribution of Perseus arm masers in projection onto the Galactic plane and the model spiral constructed for them by the three-point method at  $(\Delta l)_{\text {min}}^0 =49^\circ$. The error bars reflect the parallax uncertainty. The
dotted lines mark the boundaries of the $1\sigma$ confidence
region for the model spiral.
}
\label{Per_conf_reg}
\end{figure}

The results of our analysis of the positions of
Perseus arm masers for various values of $(\Delta l)_{\text {min}}$
are summarized in Table~\ref{Per_tab}. These results show that
as~$(\Delta l)_{\text {min}}$ increases, the error of the mean~$\sigma_{\langle R_0 \rangle}$
and the confidence interval for the median of $R_{0,j}$
initially decrease because of the truncation of the $R_{0,j}$
distribution wings (Fig.~\ref{f_Per}) and, once the variance of
this distribution has achieved stabilization (see the
row of standard deviations $\sigma_{R_0}$ in Table~\ref{Per_tab}), grow due
to the reduction in the sample size (the number of
solutions~$N_{\text{sol}}$). This makes it possible to choose
an optimal longitude constraint~$(\Delta l)_{\text {min}}^0$ from the
minimum of the formal error of the mean $\sigma_{\langle R_0 \rangle}$ (a
more stable dispersion characteristic than the length
of the confidence interval for the median). For the
Perseus arm we found $(\Delta l)_{\text
{min}}^0=49^\circ$. The model
spiral corresponding to this constraint in comparison
with the positions of the masers is presented in Fig.~\ref{Per_conf_reg}.

\begin{figure}[t!]
\centerline{%
\epsfxsize=17cm%
\epsffile{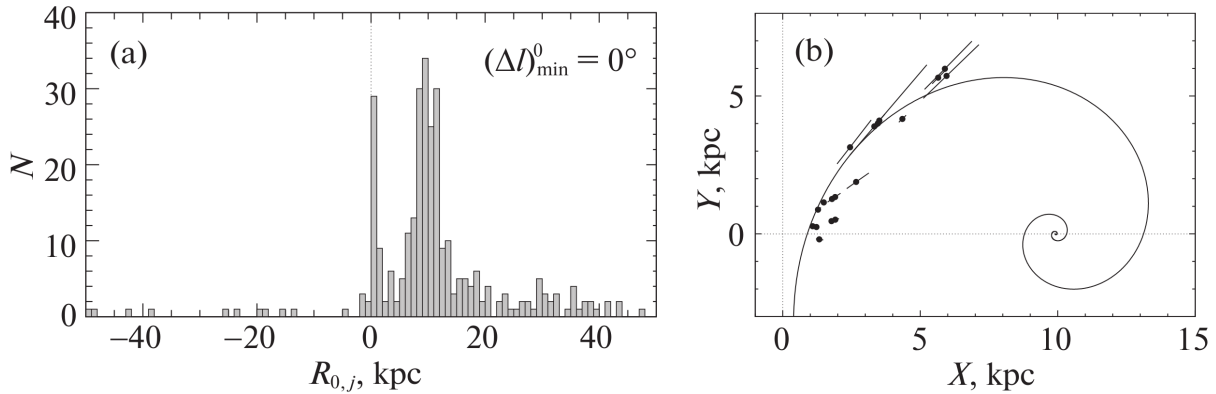}%
}
\vskip-0.3cm
\caption{\rm Visualization of the results of applying the three-point method to the Sagittarius arm at $(\Delta l)_{\text {min}}^0 =0^\circ$. (a) The $R_{0,j}$\,
distribution. (b) The model spiral in comparison with the positions of the masers assigned to this arm in projection onto the
Galactic plane (the error bars reflect the parallax uncertainty).
}
\label{f_Sgr}
\end{figure}

\begin{figure}[t!]
\centerline{%
\epsfxsize=17cm%
\epsffile{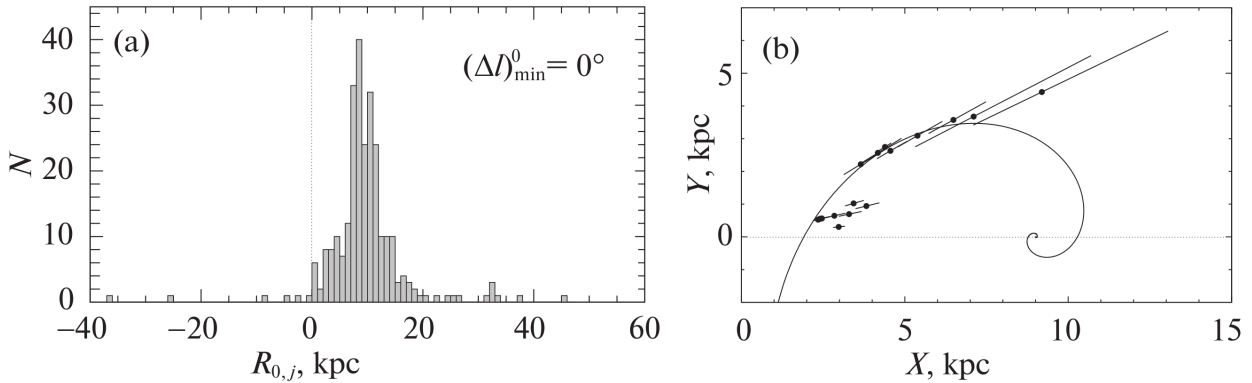}%
}
\vskip-0.3cm
\caption{\rm Same as Fig.~\ref{f_Sgr} for the Scutum arm; $(\Delta l)_{\text {min}}^0 =0^\circ$.}
\label{f_Sct}
\end{figure}

\begin{figure}[t!]
\centerline{%
\epsfxsize=17cm%
\epsffile{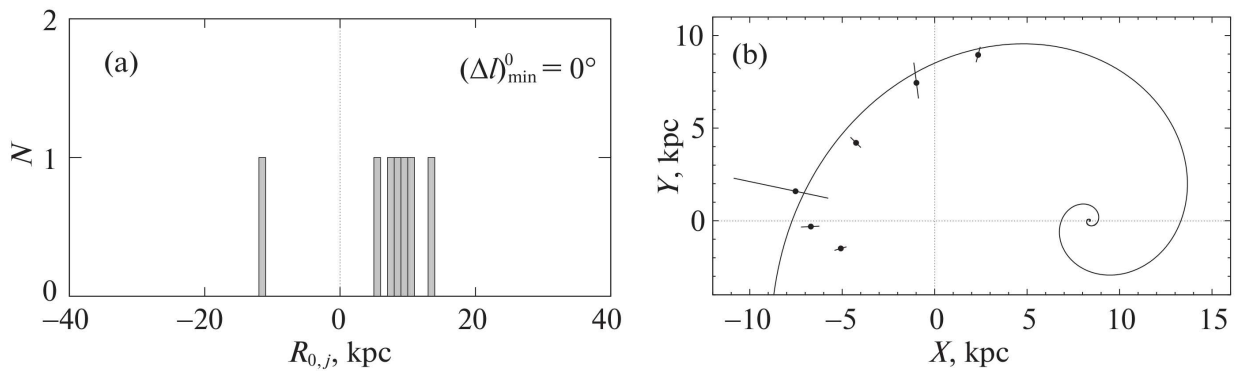}%
}
\vskip-0.3cm
\caption{\rm Same as Fig.~\ref{f_Sgr} for the Outer arm; $(\Delta l)_{\text {min}}^0 =0^\circ$.}
\label{f_Out}
\end{figure}

The parameters of other spiral segments were determined in a similar way. All masers of the Scutum
arm and almost all masers of the Sagittarius arm are
on one side of the $X$ axis (Figs.~\ref{f_Sgr}b and~\ref{f_Sct}b), and,
consequently, all triplets of masers in the first case and
almost all of them in the second one give two solutions for each $R_0$ (see Subsection~2.2). Excluding the
triplets with a nonunique solution is then inapplicable
for these segments. Therefore, we used an alternative
processing: for each such triplet we chose the root for
which the sum of the squares of the distances from
the spiral corresponding to it to the remaining masers
of the segment under consideration was smaller. The
Outer arm was processed in the same way, because
it is represented by a small number of objects ($N=6$). The $R_{0,j}$ distributions and the model spirals
constructed for these three segments at $(\Delta l)_{\text
{min}}^0 =0^\circ$ in comparison with the positions of the masers are
shown in Figs.~\ref{f_Sgr}--\ref{f_Out}.

\begin{table}[t!]
 \centering
 \tabcolsep 0.105cm
 \caption
 {\label{Per_tab} \rm Results of applying the three-point method to the Perseus arm masers
}
\vspace{5mm}
 \begin{tabular}{c|c|c|c|c|c|c|c|c|c}
   \hline
 $(\Delta l)_{\text{min}}$ &$    0 ^\circ$  &$    5 ^\circ$  &$  15 ^\circ$ &$   30 ^\circ$ &$   40 ^\circ$ &$   49 ^\circ$ &$   50 ^\circ$ &$   60 ^\circ$ &$70^\circ$ \\[0.07cm]
 \hline
 $N_{\text{sol}}$  &$  462 $   &$  445 $   &$  335 $ &$  237 $ &$  213 $ &$  161 $ &$  159 $ &$  104 $ &$   34 $ \\[0.07cm]
 \hline
 $\langle{R}_0\rangle$  &$8.79$ &$9.26$ &$9.48$&$8.47 $&$7.75$&$8.55$&$8.55$&$8.54$&$9.07$ \\[0.07cm]
 $\sigma_{{R}_0}$&$8.74$ &$8.55$ &$7.04$&$5.28$&$3.52$&$1.51$&$1.52$&$1.51$&$1.48$\\[0.07cm]
 $\sigma_{\langle{R}_0\rangle}$&$0.41$ &$0.41$ &$0.38$&$0.34$&$0.24$&$0.119$&$0.121$&$0.15$&$0.25$\\[0.07cm]
 \hline
 $\me R_0$&$\bf{8.11}$ &$\bf{8.25}$ &$\bf{8.46}$&$\bf{8.23}$&$\bf{8.20}$&$\bf{8.43}$&$\bf{8.41}$&$\bf{8.42}$&$\bf{8.92}$\\[0.07cm]
 $\sigma^+(\me R_0)$&$+0.22$ &$+0.21$ &$+0.29$&$+0.20$&$+0.17$&$+0.19$&$+0.21$&$+0.35$&$+0.69$\\[0.07cm]
 $\sigma^-(\me R_0)$&$-0.22$ &$-0.24$ &$-0.23$&$-0.26$&$-0.30$&$-0.20$&$-0.18$&$-0.20$&$-0.55$\\[0.07cm]
   \hline
  $\me i$     &$ \bf{-9 \fdg69 } $   &$ \bf{-9 \fdg85 } $   &$ \bf{-9 \fdg83 } $ &$ \bf{-9 \fdg76 } $ &$ \bf{-9 \fdg75 } $ &$ \bf{         -10 \fdg61 } $ &$ \bf{         -10 \fdg43 }$ &$ \bf{         -10 \fdg61 } $ &$ \bf{         -12 \fdg 1} $ \\[0.07cm]
 $\sigma^+(\me i)$   &$ + 0 \fdg35 $   &$ + 0 \fdg27 $   &$ + 0 \fdg47 $ &$ + 0 \fdg91 $ &$ + 1 \fdg 0 5 $ &$ + 0 \fdg75 $ &$ + 0 \fdg57 $ &$ + 0 \fdg64 $ &$ + 1 \fdg3 $ \\[0.07cm]
 $\sigma^-(\me i)$    &$ - 0 \fdg23 $   &$ - 0 \fdg19 $  &$ - 0 \fdg     22      $ &$ - 0 \fdg64 $ &$ - 0 \fdg65 $ &$ - 0 \fdg64 $ &$ - 0 \fdg95 $ &$ - 1 \fdg54 $ &$ - 3 \fdg7 $ \\[0.07cm]
 \hline
 $\me \lambda_0$    &$ \bf{78 \fdg2 } $  &$ \bf{78 \fdg7 } $   &$ \bf{73 \fdg 0 } $ &$ \bf{73 \fdg 0  } $ &$ \bf{78 \fdg4 } $ &$ \bf{60 \fdg7 } $ &$ \bf{61 \fdg4 } $ &$ \bf{61 \fdg1 } $ &$ \bf  {      52 \fdg 1 } $ \\[0.07cm]
 $\sigma^+(\me \lambda_0)$      &$ + 7 \fdg6 $   &$ + 8 \fdg5 $  &$ + 6 \fdg 1 $ &$ + 6 \fdg8 $ &$ +10 \fdg5 $ &$ +10 \fdg 0  $ &$ + 9 \fdg5 $ &$ + 9 \fdg8 $ &$ +10 \fdg6 $ \\[0.07cm]
 $\sigma^-(\me \lambda_0)$      &$ - 7 \fdg3 $   &$ - 5 \fdg7 $   &$ - 8 \fdg6 $ &$ - 9 \fdg8 $ &$ -13 \fdg3 $ &$ - 2 \fdg8 $ &$ - 3 \fdg3$ &$ - 2 \fdg9 $ &$ -10 \fdg4 $ \\[0.07cm]
\hline
 $(\sigma_{\text w})_{\varpi}$       &$    0.23 $  &$    0.23 $  &$    0.23 $ &$    0.24 $ &$    0.23 $ &$    0.25 $ &$    0.25 $ &$    0.25 $ &$    0.24 $ \\[0.07cm]
 $(\sigma_{\text w})_{\text {obs}}$  &$    0.44 $   &$    0.48 $  &$    0.44 $ &$    0.43 $ &$    0.45 $ &$    0.46 $ &$    0.46 $ &$    0.45 $ &$    0.44 $ \\[0.07cm]
  $(\sigma_{\text w})_0$             &$    0.37 $ &$    0.42 $   &$    0.37 $ &$    0.35 $ &$    0.39 $ &$    0.38 $ &$    0.39 $ &$    0.38 $ &$    0.37 $ \\[0.07cm]
 \hline
\multicolumn{10}{l}{}\\ [-3mm]
\multicolumn{10}{p{15.5cm}}{{\footnotesize$\langle R_0\rangle$ and $\sigma_{\langle R_0
\rangle}$ are the arithmetic mean of $N_{\text{sol}}$ values of $R_{0,j}$ and its error; $\sigma_{R_0}$ is the standard deviation of the $R_{0,j}$ distribution; $\sigma^{+}$ 
and $\sigma^{-}$ are the statistical uncertainties of the median for the $1\sigma$ confidence level toward larger and smaller values, respectively; $(\sigma_{\text w})_{\varpi}$
is the contribution to the scatter of masers across the segment from the parallax uncertainty; $(\sigma_{\text w})_{\text {obs}}$ and $(\sigma_{\text w})_0$ are, respectively, the
observed and natural rms scatters across the segment.$R_0$, $(\sigma_{\text w})_{\varpi}$, $(\sigma_{\text w})_{\text
{obs}}$, and $(\sigma_{\text w})_0$ are given in kiloparsecs.

}}
 \end{tabular}
 \end{table}
 
\begin{figure}[t!]
\centerline{%
\epsfxsize=17cm%
\epsffile{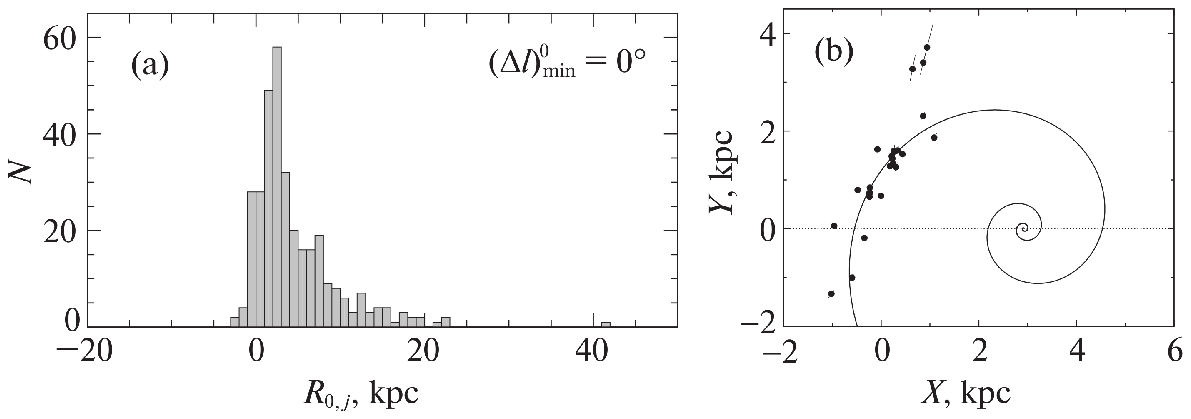}%
}
\vskip-0.3cm
\caption{\rm Same as Fig.~\ref{f_Sgr} for the Local arm. The triplets with multiple roots of the equation for $R_0$ were excluded from
consideration; $(\Delta l)_{\text {min}}^0 =0^\circ$.}
\label{f_Loc}
\end{figure}

\begin{figure}[t!]
\centerline{%
\epsfxsize=17cm%
\epsffile{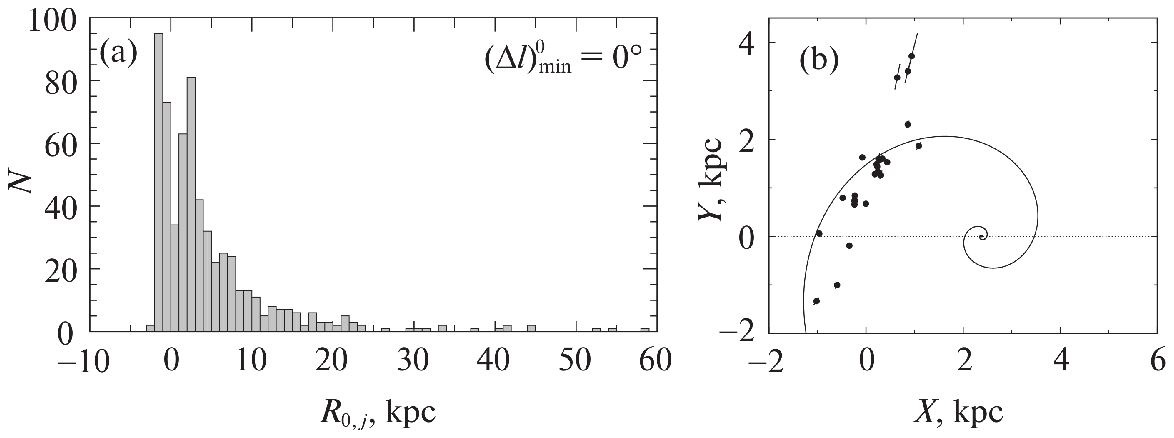}%
}
\vskip-0.3cm
\caption{\rm  Same as Fig.~\ref{f_Loc} when choosing the solution with the smallest scatter of masers relative to the model spiral from the
multiple roots; $(\Delta l)_{\text {min}}^0 =0^\circ$.}
\label{f_Loc0}
\end{figure}

For the Local arm we considered both techniques
of processing the triplets with a nonunique solution
for $R_0$. A visualization of the results obtained is
presented in Figs.~\ref{f_Loc} and~\ref{f_Loc0}. We took the results
obtained by excluding the triplets with a nonunique
solution as the final ones for this segment, because
the model spiral describes better the positions of the
segment masers in this case (cf. Figs.~\ref{f_Loc}b,~\ref{f_Loc0}b).

A summary of the main results for the individual
Galactic spiral arm segments is given in Table~\ref{tab_all}. The
absence of~$(\sigma_{\text{w}})_0$  for the Outer and Scutum arms in
the last column of the table corresponds to a negative
estimate of the natural variance across the arm~$(\sigma_{\text{w}})_0^2$.
This means that the observed scatter of masers relative to these segments is entirely (and even excessively) explained by the catalogue uncertainties of the
trigonometric parallaxes in Reid et al.\ (2014) (see
Section~5).

Since the results for the individual segments were
obtained from small samples of objects, it is important
to estimate the biases of $R_0$ and other parameters
found by the three-point method due to the sample
finiteness. This was done using the jackknife technique (see Appendix~A3) that also allows the variances of parameters to be estimated. The corrected 
results are presented in Table~\ref{tab_all_JK}. Comparison with 
Table~\ref{tab_all} shows that the jackknife uncertainties of the 
parameters turned out to be appreciably larger than 
the formal uncertainties of the median estimates in 
the overwhelming majority of cases. Obviously, this 
is explained by the fact that each sample object, as a 
rule, enters into different triplets, and this makes the 
  results for individual triplets not quite independent. 
  The latter is naturally taken into account when using 
  the jackknife technique. Our numerical experiments 
  (see Section~\ref{MonteCarlo}) confirm that the jackknife estimates of 
    the variances are more adequate. For completeness, 
    we provide both error estimates. For practical purposes, we will choose the larger of them. 

\begin{table}[t!]
  \centering
    \caption{\label{tab_all}\rm Results of applying the three-point method to five spiral arm segments for optimal longitude constraints~$(\Delta
    l)_{\text {min}}^0$}
\vspace{5 pt}
\tabcolsep0.35cm
       \begin{tabular}{c|c|c|c|c|l|c}
   \hline
   Arm & $N_{\text{sol}}$ &  $(\Delta l)_{\text{min}}^0$ &  $\me R_0,~$kpc& $ \me i $ & $\quad \me \lambda_0$ & $(\sigma_{\text{w}})_0$, kpc \\
 \hline
  {Sct} & 267 & $0^\circ$ & $9.01^{+0.30}_{-0.15}$ & $-28\fdg64^{+1\fdg40^{\vphantom{T^T}}}_{-1\fdg59}$ & $-25\fdg0^{+3\fdg5}_{-4\fdg5}$\\[0.2cm]
  Sgr & 306 & $0^\circ$ & $9.92^{+0.36}_{-0.34}$ & $-18\fdg29^{+0\fdg99}_{-1\fdg13}$ &  $-16\fdg4^{+2\fdg4}_{-3\fdg9}$  & $0.34$\\[0.2cm]  
   Loc & $328$ & $0^\circ$ & $2.93^{+0.35}_{-0.19}$ & $-13\fdg85^{+0\fdg39}_{-0\fdg37}$ & $+39\fdg7^{+6\fdg7}_{-4\fdg9}$ & $0.52$\\[0.2cm]
  {Per} & 161 &$49^\circ$ & $8.43^{+0.19}_{-0.20}$ & $-10\fdg61^{+0\fdg75}_{-0\fdg64}$ & $+60\fdg7^{+10\fdg0}_{-2\fdg8}$ & $0.38$\\[0.2cm]
 Out & 7 & $0^\circ$ &  $8.4^{+5.3}_{-19.6}$ & $-20\fdg6^{+16\fdg9}_{-58\deg}$ & ${+100^\circ}^{+521^\circ}_{-56^\circ}$ \\[0.2cm]
   \hline
  \end{tabular}
\end{table}

\begin{table}[t!]
  \centering
    \caption{\label{tab_all_JK}\rm Results of correcting the estimates of the spiral segment parameters by the jackknife technique }
    \vspace{5 pt}
    \tabcolsep0.22cm
  \begin{tabular}{l|c|c||c|c||l|c}
  \hline
   Arm & $R_{0,\text{corr}} \pm \sigma_{R_0,\text{J}},~$kpc & $\Delta R_0,~$kpc & $i_{\text{corr}} \pm \sigma_{i,\text{J}}$ & $\Delta i$ & $\lambda_{0,\text{corr}} \pm \sigma_{\lambda_0,\text{J}}$ & $\Delta \lambda_0$ \\
   \hline
     Sct &  $8.62 \pm 0.81$ & $-0.39$  & $-27\fdg2 \pm 5\fdg4$ & $+1\fdg4$ & $-21\fdg1 \pm 12\fdg0$ & $+3\fdg9$ \\[0.2cm]
  Sgr &  $10.62 \pm 0.69$  & $+0.70$ & $-15\fdg5 \pm 3\fdg6$ & $+2\fdg8$  & $-12\fdg8 \pm 8\fdg7$ & $+3\fdg6$ \\[0.2cm]
   Loc & $2.17 \pm 1.00$ & $-0.76$  & $-12\fdg3 \pm 4\fdg6$ & $+1\fdg6$  & $+39\fdg1 \pm 18\fdg8$ & $-0\fdg6$ \\[0.2cm]
  Per &   $8.36 \pm 0.53$ & $-0.07$  & $-11\fdg5 \pm 2\fdg2$ & $-0\fdg9 $  & $+36\fdg9 \pm 15\fdg0$ & $-23\fdg8 $ \\[0.2cm]
   Out &   $16.0 \pm 7.5 $ & $+7.6 $  & $+3\fdg5 \pm 25\deg$ & $ +24\deg$ & $+84\deg \pm 40\deg$ & $-16\deg $\\[0.07cm]
 \hline
\multicolumn{7}{l}{}\\ [-3mm]
\multicolumn{7}{p{15.5cm}}
{{\footnotesize $R_{0,\text{corr}}$,  $i_{\text{corr}}$, and 
$\lambda_{0,\text{corr}}$ are the corrected estimates of the parameters;  $\sigma_{R_0,\text{J}}$,
$\sigma_{i,\text{J}}$, and $\sigma_{\lambda_0,\text{J}}$ are the jackknife estimates of the
uncertainties in the parameters; $\Delta R_0$, $\Delta i$, and $\Delta \lambda_0$ are the jackknife corrections to the estimates of the parameters (the differences
between the corrected estimates and the median values in Table~\ref{tab_all}).

}}
\end{tabular}
\end{table}

\subsection{The Mean Estimate of $R_0$ from the Results 
of Applying \\ the Three-Point Method to Masers }
\label{meanR0}

Table~\ref{tab_all_JK} shows that the results for different segments differ sharply in solution reliability. The Outer
arm has the worst characteristics (the largest biases
and variances of the parameters), because the sample
of masers is very small in size ($N = 6$). In fact,
this segment creates no significant constraints on $R_0$,
 and, therefore, the result for it was not used in deducing the mean estimate of $R_0$. A similar decision was
made with regard to the Sagittarius and Local arms.
For the Sagittarius arm the $R_{0,j}$ distribution function
is definitely bimodal (Fig.~\ref{f_Sgr}a), suggesting that the
 solution is not unique; in that case, even the median
estimate is unreliable (this can be seen from Fig.~\ref{f_Sgr}b).
The model spirals for the Local arm in both versions
of the three-point method agree poorly with the positions of the masers with the largest Galactocentric
longitude (Figs.~\ref{f_Loc}b and~\ref{f_Loc0}b) that, in principle, are
most valuable for localizing the position of the spiral
segment pole, which led to an inconsistent solution.

The results for the Perseus and Scutum arms are
much more reliable: the estimates of $R_0$ from them
have the smallest biases and acceptable variances.
Obviously, this is explained predominantly by the fact
that the Perseus arm segment has the greatest extent
(in both $l$ and $\Lambda$) and contains relatively many masers,
while the Scutum arm (with a moderate number of
masers) is innermost and, hence, determines better
the position of the spiral pole, other things being equal
(Table~\ref{tab_masers}, Fig.~\ref{pic_88}).

For these reasons, we found the mean estimate
$\langle R_0\rangle$  by this method based on the corrected values
of $R_0$ for the Perseus and Scutum arms given in
Table~\ref{tab_all_JK} as a weighted mean with weights inversely
proportional to the variances  $\sigma_{R_0,\text{J}}^2$ determined by the
jackknife technique:
\begin{equation}
\begin{split}
 \langle R_0\rangle &= \left(8.36/0.53^2 +   8.62/0.81^2\right)\left/
    \left(1/0.53^2 +  1/0.81^2\right)\right. = 8.44~\text{kpc}, \\
 \sigma_{\langle R_0\rangle} &= \left(1/0.53^2 + 1/0.81^2\right)^{-1/2} = 0.45~\text{kpc}.
\end{split}
\end{equation}

At fixed $R_0$ the three-point method becomes the
two-point method of determining the segment pitch
angle $i$ and position parameter $\lambda_0$. In this way we
estimated $i$ and $\lambda_0$ for all five segments at $R_0=8.44$~kpc: for each segment these parameters were
determined as the medians of the values obtained
from all possible pairs of segment masers with the
minimum distance between the pair objects $(\Delta l)^0_{\text{min}}$
determined by applying the three-point method to this
segment. The results are presented in Table~\ref{tab_8.44}.

\begin{table}[t!]
   \centering
     \caption{\label{tab_8.44}\rm      Results of determining the parameters $i$ and $\lambda_0$ and dispersion characteristics of the spiral arm segments at fixed $R_0 = 8.44~$kpc}
\vspace{5 pt}
\tabcolsep0.33cm
     \begin{tabular}{c|cc|cc|cc}
 \hline
  Arm  & $\me i$ & $ {\sigma}_{i,\text{J}}$ & $\me \lambda_0$ & ${\sigma}_{\lambda_0,\text{J}}$ & $(\sigma_{\text{w}})_{\text{obs}}\,,~$kpc   & $(\sigma_{\text{w}})_{0}\,,~$kpc \\
 \hline
  Sct   & ${-21\fdg4}^{+0\fdg6}_{-1\fdg0}$  & ${1\fdg8}$ & $-43\fdg9^{+2\fdg8^{\vphantom{T^T}}}_{-5\fdg7}$ & $ 10\fdg5$ & $0.51 \pm 0.26$ & $ $\\ [0.2cm]
Sgr  & ${-9\fdg9}^{+1\fdg8}_{-0\fdg8}$ & $ {3\fdg6}$ & $-50\fdg8^{+6\fdg5}_{-16\fdg7}$ & $26^\circ$ & $0.37 \pm 0.08$ & $ {0.20} \pm {0.04}$ \\ [0.2cm]
 Loc & ${-16\fdg5}^{+1\fdg4}_{-2\fdg2}$ & $ {5\fdg1} $ & $+9\fdg0^{+0\fdg3}_{-0\fdg2}$ & $ 0\fdg6$ & $0.30 \pm 0.06$  & ${0.29} \pm {0.04}$ \\[0.2cm]
 Per   & ${-10\fdg6}^{+0\fdg6}_{-0\fdg4}$  & ${1\fdg1}$ & $+63\fdg3^{+4\fdg3}_{-2\fdg1}$ & $ 9\fdg4$  & $0.42 \pm0.08$  & ${0.34} \pm {0.05}$  \\ [0.2cm]
 Out   & ${-18\fdg6}^{+6\fdg7}_{-5\fdg6}$ & ${ 0\fdg8}$ & $+98\deg^{+26\deg}_{-11\deg}$ & $ 2\fdg0$ & $1.19 \pm 0.49$  & $ $ \\ [0.2cm]
 \hline
\multicolumn{7}{l}{}\\ [-3mm]
\multicolumn{7}{p{15.5cm}}
{\footnotesize $\sigma_{i,\text{J}}$ and $ \sigma_{\lambda_0,\text{J}}$ are the jackknife estimates of the uncertainties in the median values.

}
 \end{tabular}
\end{table}

\begin{table}[t!]
 \centering
 \caption{\label{tab_8.8} \rm Results of determining the parameters $i$ and $\lambda_0$ and dispersion characteristics of the spiral arm segments at the
final $R_0 = 8.8~$kpc}
\vspace{5 pt}
\tabcolsep0.4cm
 \begin{tabular}{c|cc|cc|cc}
 \hline
  Arm  & $\me i$ & $ {\sigma}_{i,\text{J}}$ & $\me \lambda_0$ & ${\sigma}_{\lambda_0,\text{J}}$ & $(\sigma_{\text{w}})_{\text{obs}}\,,~$kpc  & $(\sigma_{\text{w}})_{0}\,,~$kpc \\
 \hline
  Sct   & $-23\fdg7_{-3\fdg7}^{+1\fdg1} $  & $1\fdg1 $ & $-36\fdg9_{-6\fdg5}^{+3\fdg1^{\vphantom{T^T}}} $ & $5\fdg7 $ & $0.48 \pm 0.23$ & $ $\\ [0.2cm]
 Sgr  & $-9\fdg9_{-2\fdg0}^{+0\fdg5} $ & $  3\fdg1$ & $ -44\fdg4_{-9\fdg7}^{+3\fdg2}$ & $22\deg $ & $0.39 \pm 0.09$ & $ 0.29\pm0.08 $ \\ [0.2cm]
 Loc & $ -16\fdg8^{+1\fdg2}_{-1\fdg3}$ & $ 5\fdg4  $ & $+8\fdg5_{-0\fdg8}^{+0\fdg2} $ & $  0\fdg5$ & $0.30 \pm 0.06$  & $0.30\pm0.05 $ \\[0.2cm]
  Per   & $ -11\fdg8^{+0\fdg7}_{-0\fdg4}$  & $1\fdg5 $ & $+57\fdg2^{+5\fdg3}_{-1\fdg2} $ & $3\fdg8 $ & $0.42 \pm 0.10$   & $ 0.35\pm0.05$  \\ [0.2cm]
 Out   & $-19\fdg2_{-2\fdg9}^{+2\fdg5} $ & $0\fdg9 $ & $ +91\fdg6^{+18\fdg1}_{-5\fdg3} $ & $2\fdg3 $ & $1.19 \pm 0.48$  & $ $ \\ [0.2cm]
  \hline
\multicolumn{7}{l}{}\\ [-3mm]
\multicolumn{7}{p{15cm}}
{{\footnotesize $\sigma_{i,\text{J}}$ and $ \sigma_{\lambda_0,\text{J}}$ are the jackknife estimates of the uncertainties in the median values.

}}
 \end{tabular}
\end{table}

\section{INVESTIGATION OF THE THREE-POINT
ALGORITHM \\ BY THE MONTE CARLO
TECHNIQUE \\ AND THE FINAL SOLUTION
FROM MASERS}
\label{MonteCarlo}
As the models of spiral arms we considered logarithmic spirals with parameters $i$, $\lambda_0$, and $(\sigma_{\text{w}})_{\text{\,obs}}$ 
representing the Perseus and Scutum arms under the
assumption of $R_0 = 8.44$~kpc (Table~\ref{tab_8.44}).

The pseudo-random catalogues of objects were
generated as follows. The position of each object $O_j$
was shifted relative to the model point $M_j$, which is
an orthogonal projection of the nominal position (according to the initial catalogue) of the $j$th object onto
the model spiral; the shift was done along a straight
line perpendicular to the model spiral at point $M_j$.
The distance $\rho_j\equiv|O_jM_j|$ was varied according to
a normal law with zero mean and a standard deviation $(\sigma_{\text{w}})_{\text{\,obs}}$\,. For each of the two segments we
produced $n_{\text{MC}} = 1000$ catalogues.

For each pseudo-random catalogue the spiral segment parameters were determined by the three-point
method, with the minimum distance in Galactic longitude~$(\Delta l)_{\text{min}}$ between the adjacent points ("objects") in the set of three points having been optimized
in the same way as for the real data; i.e., we took~$(\Delta l)_{\text{min}}$ at which the statistical error of the mean for
$R_0$ calculated from all of the suitable triplets was minimal as the best value. The median values of the parameters with the scatter~$(\sigma_{\text{w}})_{\text{\,obs}}$ calculated for them
at optimal~$(\Delta l)_{\text{min}}$ were considered as a solution for
each catalogue. Our processing of $n_{\text{MC}}$ catalogues
showed that $\me R_0 - R_0=-0.66\pm 0.04$~kpc for the
Perseus arm and $\me R_0 - R_0 =
-0.08\pm 0.03$~kpc for the Scutum arm.

The $R_{0,\text{corr}}$ estimates obtained in Subsection~\ref{meanR0}
for the Perseus and Scutum arms (Table~\ref{tab_all_JK}) were
corrected for the systematic error found by the Monte
Carlo technique, which led to the point estimates
$R_0 = 9.02$ and $R_0 = 8.70$~kpc for the Perseus and Scutum
arms, respectively. As the final estimate obtained by
applying the three-point method to masers we took
the weighted mean of these two values:
\begin{equation}
 \langle R_0\rangle = \left({w_1\cdot 9.02 + w_2\cdot 8.70}\right)\left/
 \left(w_1+w_2\right)\right. = 8.8~\text{kpc}.
\end{equation}
Here, the weights $w_1=1/(1.50^{2}+0.08^2)$ and $w_2=1/(1.32^2+0.06^2)$
 take into account the lengths of the
confidence intervals for the two initial estimates (the
boundaries of the intervals were determined as the
order statistics of the set of values obtained by the
Monte Carlo technique for $N_{\text{MC}}=1000$ catalogues)
and their biases. The uncertainty of the final estimate
was found as the mean error of the weighted mean
before adjustment (Agekyan~1972):
\begin{equation}
 \sigma_{\langle R_0\rangle} = \left(2\sqrt{w_1+w_2}\,\right)^{-1} = 
 0.5~\text{kpc.}
 \end{equation}

 The parameters of the spiral arm segments were
estimated by the two-point method at fixed $R_0 =
8.8$~kpc (Table~\ref{tab_8.8}). To construct the model spiral
approximating the nominal distribution of masers, we
applied no jackknife corrections here. The difference in $i$, $\lambda_0$, and  $(\sigma_{\text{w}})_{\text{obs}}$  at $R_0 =
8.8$ and 8.44~kpc
(Table~\ref{tab_8.44}) is attributable to the correlation of these
parameters with $R_0$ (Table~\ref{correl_8.44}). Table~\ref{correl_8.44} gives the
linear correlation coefficients $\kappa(p_1,p_2)$ for the Perseus
and Scutum arms and the probability $P_\kappa\equiv P(|\varkappa| > |\kappa(p_1,p_2)|)$ to obtain a correlation coefficient greater
in absolute value than the measured  $|\kappa(p_1,p_2)|$ in the
absence of a correlation between the random variables $p_1$ and $p_2$ (Press et~al. 1997). These quantities
were determined through Monte Carlo simulations
for the above model spirals at $R_0 = 8.44$~kpc. All
probabilities $P_\kappa$ were found to be less than $0.05$, i.e.,
all coefficients~$\kappa$ in Table~\ref{correl_8.44} are significant. In most
cases, the correlation between the parameters is moderate, except for some pairs of parameters including~$\lambda_0$\,.

\begin{table}[t!]
  \centering
       \caption{\label{correl_8.44}\rm  Linear correlation coefficients  $\kappa(p_1,p_2)$ for the spiral segment parameters derived by the Monte Carlo technique
for the Perseus and Scutum arms }
\vspace{5 pt}
\tabcolsep0.2cm
  \begin{tabular}{l|c|c|c|c|c|c}
   \hline
  $(p_1,p_2)$ & $(R_0,i)$ & $(R_0, \lambda_0)$ & 
  $(R_0^{\vphantom{T^T}}, (\sigma_{\text {w}})_{\text{obs}})$ & 
  $(i, \lambda_0)$ & $(i^{\vphantom{T^T}}, (\sigma_{\text {w}})_{\text{obs}})$  & 
  $(\lambda_0, (\sigma_{\text {w}})_{\text{obs}})$ \\[0.05cm]
   \hline
    \multicolumn{7}{c}{Perseus arm}\\
   \hline
   $\kappa(p_1,p_2)$ & $-0.421$ & $-0.738$ & $-0.545$ & $0.735$ & $ -0.063$ & $ 0.514$ \\[0.05cm]
   $P_\kappa$        & $6.41\cdot10^{-4}$ & $1.17\cdot10^{-3}$  & $7.97\cdot10^{-4}$ & $1.16\cdot10^{-3}$ & $0.0466$ & $ 7.54\cdot10^{-4}$\\[0.05cm]
   \hline
    \multicolumn{7}{c}{Scutum arm}\\
   \hline
    $\kappa(p_1,p_2)$ & $-0.469$ & $+0.437$ & $-0.300$ & $-0.843$ & $ -0.344$ & $ 0.311$\\[0.05cm]
    $P_\kappa$        & $6.97\cdot10^{-4}$ & $6.59\cdot10^{-4}$ & $5.15\cdot10^{-4}$ & $ 1.57\cdot10^{-3}$ & $5.59\cdot10^{-4}$ & $5.26\cdot10^{-4}$ \\[0.05cm]
   \hline
\multicolumn{7}{l}{}\\ [-3mm]
\multicolumn{7}{p{15cm}}
{{\footnotesize $P_\kappa\equiv
       P(|\varkappa|>|\kappa(p_1,p_2)|)$ is the probability in the case of null hypothesis to obtain a correlation coefficient κ greater in absolute value
than the measured~$|\kappa(p_1,p_2)|$ for a pair of parameters~$(p_1,p_2)$.

}}
 \end{tabular}
 \end{table} 

\begin{figure}[t!]
\centerline{%
\epsfxsize=16cm%
\epsffile{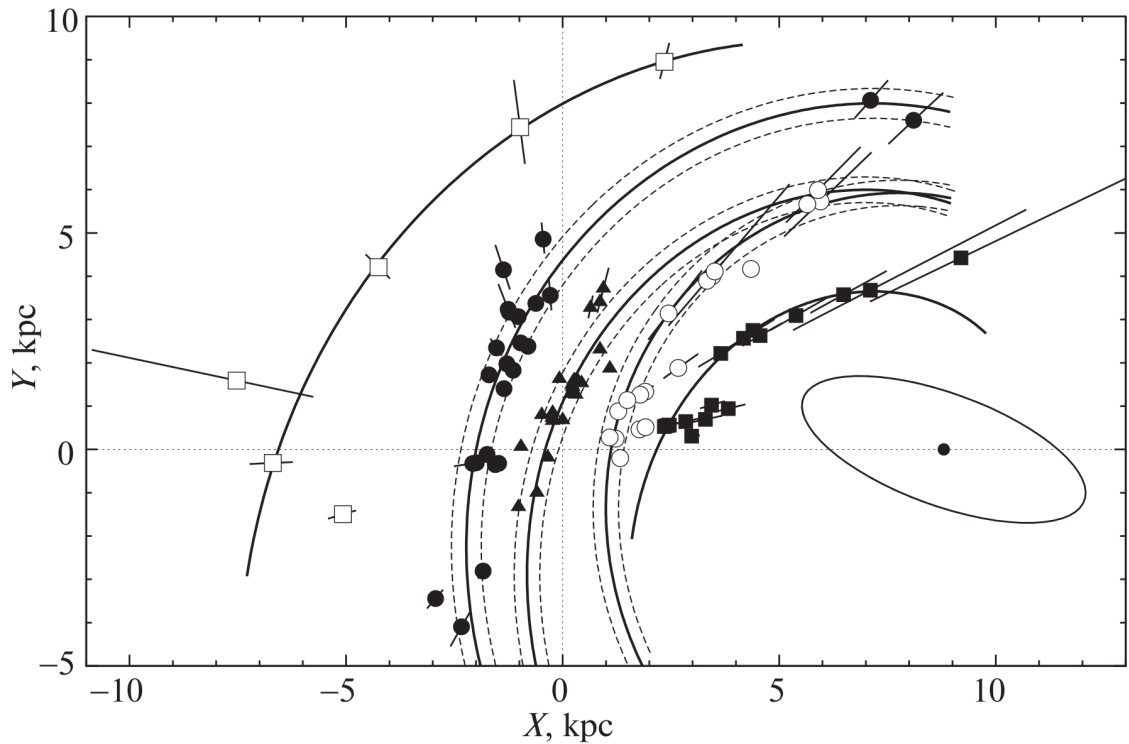}%
}
\vskip-0.3cm
\caption{\rm The model spirals (solid lines) and the maser positions in projection onto the Galactic plane for the Outer (open
squares), Perseus (filled circles), Local (filled triangles), Sagittarius (open circles), and Scutum (filled squares) arms at
$R_0=8.8$~kpc. The error bars reflect the parallax uncertainty. The natural scatter across the arms [$\pm(\sigma_{\text{w}})_{0}$ relative to the
models] is indicated by the dashed lines for segments with $(\sigma_{\text{w}})_{0}^2>0$. The ellipse indicates the Galactic bar; the circle at its center indicates the Galactic center.}
\label{pic_88}
\end{figure}

For all five segments Fig.~\ref{pic_88} presents the model
spirals with the parameters from Table~\ref{tab_8.8} and the
distribution of masers assigned to these segments in
projection onto the Galactic plane. The contours of
the Galactic bar in Fig.~\ref{pic_88} are given for an ellipsoidal
model with semiaxes $3.14:1.178: 0.81$~kpc and a
position angle $\varphi = 20^\circ$ (Casetti-Dinescu et al.\ 2013;
J\'ilkov\'a et al.\ 2012).

\section{DISCUSSION}

The three-point method is similar in principle of
generalization of {\it particular}\/ solutions for $R_0$, in our
case, obtained from triplets of objects, to the method 
of Feast and Shuttleworth (1965) within a different, 
kinematic, approach to determining $R_0$. The Feast--Shuttleworth method, which suggests the construction and analysis of the distribution function of individual $R_0$ estimates based on {\it individual}\/ objects, 
was previously quite popular (Balona and Feast 1974; 
Crampton et al.\ 1976; Loktin 1979). Such simplified 
methods are less efficient in the probabilistic sense 
than the simultaneous solution of the parameter optimization problem from all sample objects, but they 
allow one to show easily and clearly that the approach 
is operable and to judge the existence of a solution and 
its quality. Therefore, applying such methods is quite 
justified at a certain stage of using a particular approach, especially if it is new. For example, in the next 
paper we are going to investigate some properties of 
the estimates for the distances to the pole of spiral 
arms by the three-point method. However, a method 
based on the search for an extremum of the target 
function should, of course, be developed in future for 
the proposed approach. 

Note that the least-squares method (LSM) cannot 
be such a method, because it is not proper enough for 
this problem. For example, in the popular case of applying the LSM for optimization in the $(\Lambda,\ln
R/R_0)$ plane, none of the coordinates of the latter is a directly 
 measurable quantity or argument, because the heliocentric distance characteristic, in the case of masers, 
 this is the trigonometric parallax $\varpi$, is a measurable 
 (with an error) quantity ($l$ and $b$ in this problem may 
 be deemed errorless). The value of $\varpi$
 determines at 
 once both $\ln R/R_0$  and $\Lambda$. This makes the latter two 
  quantities correlated differently, depending on the object's position. The scatter of objects across the arm also leads to a similar (also complex) correlation between  $\ln R/R_0$ and $\Lambda$. These correlations are difficult
  to take into account, if at all possible. It is reasonable
 to assume the parallaxes~$\varpi$ (they are measured) and
the deviation of an object from the spiral model in the
configuration space (in the $XY$ plane) to be normally
 distributed. There is no reason to assume $\ln R/R_0$
and $\Lambda$ to be normally distributed quantities, as suggested in the LSM, nor is there reason to consider
some of these variables to be errorless. In principle,
the LSM can be used to optimize the deviations from
the spiral model in the $XY$ plane. However, on the
one hand, this leads to a sharp complication of the
calculations and, on the other hand, this will still be
insufficient, because the LSM disregards the errors
 in the distances (in~$\varpi$). The latter cannot be ignored:
they are often significant and affect differently the
     result, depending on the object's position (Fig.~\ref{pic_88}).
      These two types of scatter can be properly taken into
     account only within the maximum likelihood method,
    although this variant is more complex and laborious.
   We are working in this direction.

$R_0= 8.8\pm 0.5$~kpc found in this paper exceeds
the resent mean ("best") estimates of this parameter $\langle R_0\rangle_{\text{best}}=(7.9\div 8.3)
\pm (0.1\div 0.4)$~kpc (Genzel
       et al.\ 2010; Foster and Cooper 2010; Nikiforov and
      Smirnova 2013; Bland-Hawthorn and Gerhard 2016;
     de Grijs and Bono 2016), but it is within the interval
    covered by the point estimates of $R_0$ in modern
   publications. For example, Do et al.\ (2013) obtained
  $R_0=8.92^{+0.58}_{-0.55}$~kpc through three-dimensional Jeans
  modeling (within the statistical parallax method) of
 the stellar kinematics of the Galactic nuclear star
cluster; Catchpole et al.\ (2016) deduced a similar
estimate, $R_0=8.9\pm {\sim} 0.4$~kpc, from Mira variables
in the bulge. Given the errors, our result from the
geometry of spiral segments is consistent with the
estimate $R_0=8.34\pm
0.16$~kpc from the kinematics of
masers based on the same database (Reid~et~al. 2014).

The $i$ estimates in Table~\ref{tab_8.8}  confirm the difference
between different spiral arms in pitch angle found by
Reid et al.\ (2014). According to our results, this
difference is significant when comparing the Scutum
and Outer arms, on the one hand, and the Sagittarius
and Perseus arms, on the other hand. An extrapolation of the model spiral of the Local arm shows
that it can be a trailing branch of the Sagittarius arm
(Fig.~\ref{pic_88}), at least on the basis of masers. This variant
was not considered in Xu et al.\ (2013) devoted to the
nature of the Local arm.

In the case of segments with the most reliable
solutions (the Perseus and Sagittarius arms), $R_0$ affects noticeably the $i$ estimate (cf. Tables~\ref{tab_8.44} and~\ref{tab_8.8}) in
accordance with the correlation coefficients (Table~\ref{correl_8.44}).
This should be kept in mind when discussing the
results of various pitch angle measurements for the
Galactic spiral arms.

The natural scatters~$(\sigma_{\text {w}})_0$  of masers across the
Sagittarius, Local, and Perseus arms were found to
be close to those found by Reid et al.\ (2014). However, for the Scutum and Outer arms, the farthest
ones from the Sun among those revealed by these
data, the variance~$(\sigma_{\text {w}})_0^2$ turned out to be negative in
all variants of calculations. This may imply that the
parallax uncertainties given by Reid et al.\ (2014) for
the masers of these two arms were overestimated. An
alternative reason can be the selection effects when
choosing the masers for the samples of these arms.
For example, the Scutum sample consists of two
groups inhomogeneous in~$(\sigma_{\text {w}})_{\text{obs}}$  (and in parallax
errors), one of which (at small $l$) lies at a large angle
to the spiral segment and has a significant scatter
relative to it, while the other group is located in a
direction almost tangential to the segment ($l\sim 30\deg$),
in a narrow interval of $l$, which leads to a scatter
across the segment much smaller than that for the
first group.

Note that the assumption (number 5 in our list
in the Introduction) about the coincidence between
the pole of the spiral arms (in a more general case,
the geometric center of the spiral pattern) and the
Galactic center is currently a standard one in the
parametric modeling of the Galactic spiral structure
(see the review of Efremov (2011) and the papers
on the spiral structure mentioned in the Introduction). At present, this assumption seems reasonable, because it is consistent with the observations
of external spiral galaxies (see, e.g., Savchenko and
Reshetnikov 2013), the numerical experiments (e.g.,
Korchagin et al.\ 2016), and the present views that
the inner Galaxy is a dynamically "relaxed" system
whose differently determined "centers" (the greatest
star density or a different central feature of the spatial distribution, the central object, the barycenter,
and others) virtually coincide (Bland-Hawthorn and
Gerhard 2016). The currently available data are not
yet accurate enough to directly establish the extent
to which these centers coincide with one another
and with the geometric center of the spiral structure
introduced by us, but such problems should be kept
in mind in future. Note separately that we do not use
the assumption that the spiral arm extends to the very
center of the Galaxy, considering the latter as the pole
of the approximating spiral of the observed segment.
The existence of a real arm only outside the Galactic
bar or even in a limited interval of radii $R$ is not a
hindrance in using this approach.

\section{CONCLUSIONS}
We proposed a new approach to the spatial modeling of Galactic spiral arm segments that includes the
determination of the distance to the pole of the spiral
structure, i.e., the distance to the Galactic center $R_0$.
To study the capabilities of this approach, we considered the problem of reconstructing the parameters
of a logarithmic spiral as a geometric figure from
points belonging to it by assuming the direction to the
spiral pole to be known and the points to represent a
segment constituting less than one spiral turn. Our
numerical--analytical study using the representative
spirals for the Perseus and Scutum arms as examples
leads us to the following conclusions.
\begin{itemize}
\item[(1)]Knowing the positions of four points of the segment uniquely solves the problem of reconstructing
the spiral arm parameters. However, this solution
cannot be used in practice: for any small change in
the positions of the points the solution ceases to exist,
because the spiral that passes through four arbitrary
points in one turn generally does not exist.
\item[(2)] If the positions of three points of the segment
are known, then the solution exists always, but it is
generally nonunique: apart from the initial $R_0$, there
can be one or two additional roots. In this case,
\begin{itemize}
\item[(a)] if the segment lies completely on one side of
the $X$ axis (the Galactic center--anticenter line) or
touches the extreme point of this axis, then there are
always two roots;
\item[(b)] if the segment crosses the $X$ axis, then the
root is usually unique, except for the segments that
lie {\it almost}\/ completely on one side of the $X$ axis (three
roots, two at the bifurcation points) and the short
segments with most of their length being at negative
longitudes $l$, $\Lambda$  (two roots);
\item[(c)] the additional roots usually differ greatly from
the initial $R_0$ (they are often negative for the Perseus
arm) and are distinguishable from it by the pitch
angle, except for the cases where the middle point
of the segment is near the traverse directions ($\Lambda_2\approx$ from $
-80^\circ$ to $-70^\circ$ and $\Lambda_2\approx$ from $+100^\circ$ to $+115^\circ$);
\item[(d)] the region of configurations of triplets of points
for which a unique solution exists is not small and
corresponds to the segments in the solar sector of
the Galaxy that are usually revealed by tracers with
reliable distances.
\end{itemize}
This, three-point, method can be applied to real
data and in numerical experiments provided that a
criterion for choosing between the roots is introduced.
\item[(3)] The segments that cross the $X$ axis but are
not centered near the traverse directions are preferred
when seeking a geometrically exact solution and,
probably, an approximation solution.
\end{itemize}

Based on the three-point method, we constructed
a simplified algorithm for determining the parameters
of a spiral segment from real objects. Applying the algorithm to the data from Reid et al.\ (2014) on masers
with trigonometric parallaxes confirmed that, on the
whole, the new approach is operable. We managed to
obtain reliable solutions for the Perseus and Scutum
arms. Averaging these results with the corrections
for the sample finiteness and the estimator bias led
to the final estimate of $R_0= 8.8\pm 0.5$ kpc from the
geometry of the spiral segments traced by masers.

We estimated the parameters of five spiral arm
segments revealed by masers by a similar, two-point
at fixed $R_0$, method. We confirmed that the pitch
angles for different spiral arms are generally different.
Our results suggest that the Local arm can be a
branch of the Sagittarius arm. We found a significant
negative correlation between the pitch angle $i$ and $R_0$.
We showed that the observed scatter of masers relative to the Outer and Scutum arms could generally
be explained by the catalogue uncertainties in the
trigonometric parallaxes.

\section*{ACKNOWLEDGMENTS}
We are grateful to the referees for their useful
remarks. This work was financially supported by the
St. Petersburg State University (grant no. 6.37.341.2015).

\section*{REFERENCES}
\begin{enumerate}

\item T. A. Agekyan, \emph{Fundamentals of the Error Theory
for Astronomers and Physicists} (Nauka, Moscow,
1972) [in Russian].

\item V. S. Avedisova, Sov. Astron. Lett. {\bf 11}, 185 (1985).
\item L. A. Balona and M. W. Feast, Mon. Not. R. Astron.
Soc. {\bf167}, 621 (1974).

\item J. Bland-Hawthorn and O. Gerhard, Ann. Rev. Astron. Astrophys. {\bf54}, 529 (2016).

\item V. V. Bobylev and A. T. Bajkova, Astron. Lett. {\bf39}, 759
(2013).
\item V. V. Bobylev and A. T. Bajkova, Mon. Not. R. Astron.
Soc. {\bf437}, 1549 (2014).
 
\item D. I. Casetti-Dinescu, T. M. Girard, L. J\'{i}lkov\'a, et al.,
Astron. J. {\bf146}, 33 (2013).
\item R. M. Catchpole, P. A. Whitelock, M. W. Feast,
S. M. G. Hughes, M. Irwin, and C.~Alard, Mon. Not.
R. Astron. Soc. {\bf455}, 2216 (2016).
\item D. Crampton, D. Bernard, B. L. Harris, and
A. D. Thackeray, Mon. Not. R. Astron. Soc. {\bf176}, 683
(1976).
\item A. K. Dambis, L. N. Berdnikov, Yu. N. Efremov,
A. Yu. Kniazev, A. S. Rastorguev, E. V. Glushkova,
V. V. Kravtsov, D. G. Turner, et al., Astron. Lett. {\bf41},
489 (2015).
\item T. M. Dame, B. G. Elmegreen, R. S. Cohen, and
P. Thaddeus, Astrophys. J. {\bf305}, 892 (1986).
\item T. Do, G. D. Martinez, S. Yelda, A. Ghez, J. Bullock,
M. Kaplinghat, J. R. Lu, A.~H.~G.~Peter, et al., Astrophys. J. {\bf779}, L6 (2013).
\item Yu. N. Efremov, Astron. Rep. {\bf55}, 108 (2011).
\item B. Efron and C. Stein, Ann. Statistics {\bf9}, 586 (1981).
\item M. W. Feast and M. Shuttleworth, Mon. Not. R.
Astron. Soc. {\bf130}, 245 (1965).
\item T. Foster and B. Cooper, ASP Conf. Ser. {\bf438}, 16
(2010).
\item C. Francis and E. Anderson, Mon. Not. R. Astron.
Soc. {\bf422}, 1283 (2012).
\item R. Genzel, F. Eisenhauer, and S. Gillessen, Rev. Mod.
Phys. {\bf82}, 3121 (2010).
\item D. A. Grabelsky, R. S. Cohen, L. Bronfman, and
P. Thaddeus, Astrophys. J. {\bf331}, 181 (1988).
\item R. de Grijs and G. Bono, Astrophys. J. Suppl. Ser.
{\bf227}, 5 (2016).
\item L. J{\'i}lkov\'a, G. Carraro, B. Jungwiert, and I. Minchev,
Astron. Astrophys. {\bf541}, A64 (2012).
\item A. I. Kobzar', \emph{Applied Mathematical Statistics for
Engineers and Scientists} (Fizmatlit, Moscow, 2006)
[in Russian].
\item V. I. Korchagin, S. A. Khoperskov, and A. V. Khoperskov, Baltic Astron. {\bf25}, 356 (2016).
\item A. V. Loktin, Sov. Astron. {\bf23}, 671 (1979).
\item I. I. Nikiforov, ASP Conf. Ser. {\bf316}, 199 (2004).
\item I. I. Nikiforov and E. E. Kazakevich, Izv. GAO {\bf219},
245 (2009).
\item I. I. Nikiforov and O. V. Smirnova, Astron. Nachr.
{\bf334}, 749 (2013).
\item I. I. Nikiforov and T. V. Shekhovtsova, in \emph{Stellar
Dynamics: From Classic to Modern, Proceedings
of the International Conference, St. Petersburg,
Russia, Aug. 21--27, 2000}, Ed. by L. P. Ossipkov
and I. I. Nikiforov (Saint Petersburg Univ. Press, St.~Petersburg, 2001), p. 28.
\item E. D. Pavlovskaya and A. A. Suchkov, Sov. Astron.
{\bf28}, 389 (1984).
\item M. Pohl, P. Englmaier, and N. Bissantz, Astrophys. J.
{\bf677}, 283 (2008).
\item M. E. Popova, Astron. Lett. {\bf32}, 244 (2006).
\item M. E. Popova and A. V. Loktin, Astron. Lett. {\bf31}, 171
(2005).
\item W. H. Press, S. A. Teukolsky, W. T. Vetterling, and
B. P. Flannery, \emph{Numerical Recipes in C} (Cambridge
Univ. Press, Cambridge, UK, 1997).
\item M. H. Quenouille, Ann. Math. Statistics {\bf20}, 355
(1949).
\item M. H. Quenouille, Biometrika {\bf43}, 353 (1956).
\item M. J. Reid, K. M. Menten, X. W. Zheng, A. Brunthaler, L. Moscadelli, Y. Xu, B.~Zhang, M. Sato, et al.,
Astrophys. J. {\bf700}, 137 (2009).
\item M. J. Reid, K. M. Menten, A. Brunthaler,
X. W. Zheng, T. M. Dame, Y. Xu, Y.~Wu, B. Zhang,
et al., Astrophys. J. {\bf783}, 130 (2014).
\item S. S. Savchenko and V. P. Reshetnikov, Mon. Not. R.
Astron. Soc. {\bf436}, 1074 (2013).
\item J. P. Vall\'ee, Astron. J. {\bf95}, 750 (1988).
\item Y. Xu, J. J. Li, M. J. Reid, K. M. Menten, X. W. Zheng,
A. Brunthaler, L. Moscadelli, T. M. Dame, et al.,
Astrophys. J. {\bf769}, 15 (2013).
\end{enumerate}

\section*{\sl\hfill APPENDIX}
\subsection*{A1. Determining the Confidence Region for the
Model Spiral \\ Found by the Three-Point Method}

\begin{figure}[b!]
\hspace{-0.1cm}
\centerline{%
\epsfxsize=7.5cm%
\epsffile{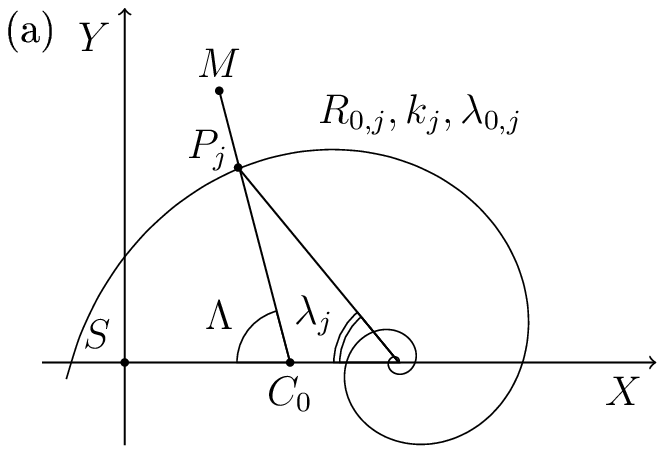}%
\raisebox{-0.7cm}{%
\hspace{0.8cm}%
\epsfxsize=7.5cm\epsffile{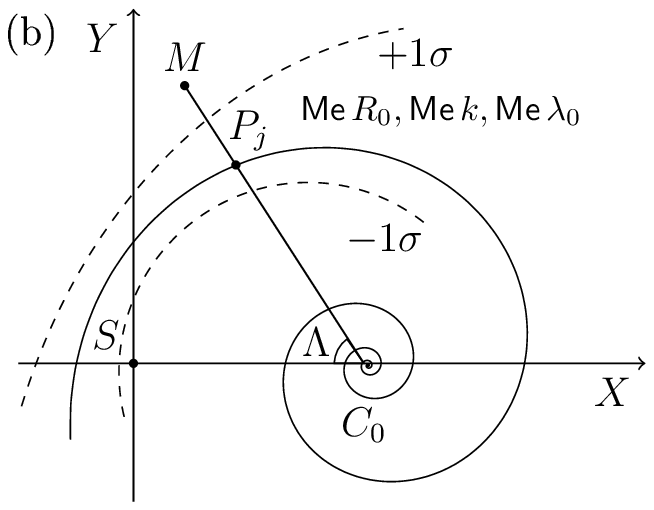}%
}%
}%
\vspace{1em}
\caption{ \rm (a) To the derivation of the equation for finding the longitude $\lambda_j$ of the point of intersection~$P_j$ of a spiral with parameters $R_{0,j}$, $k_j$, $\lambda_{0,j}$  and the ray $C_0M$. (b) To the determination of the confidence region for a model spiral with parameters $\me R_0$, $\me k$, $\me \lambda_0$\,; the dashed lines mark the boundaries of this region for the $1\sigma$; confidence level; the highlighted segment on
the ray~$C_0M$ indicates the confidence interval for a model value of~$R$ at fixed~$\Lambda$. $C_0$ is the pole of the model spiral and $S$ is the
position of the Sun.}
\label{conf_int}
\end{figure}

The~$1\sigma$ confidence region for the model spiral was
determined using the following algorithm. We fixed
the ray~$C_0M$ emerging from the pole~$C_0$ of the model
spiral at an angle~$\Lambda$ to the $X$ axis (Fig.~\ref{conf_int}a). For a
given~$\Lambda$ we determined the points of intersection $P_j$
of the ray~$C_0M$ with each of the $N_{\text{sol}}$ spirals obtained
from triplets of masers $M_{1,j}$,~$M_{2,j}$,~$M_{3,j}$\,.

Let us derive the equation to find the longitude~$\lambda_j$
of the point of intersection $P_j(X_{P_j}\,,Y_{P_j})$ of the spiral
and the ray with its origin on the $X$ axis. For point
$P_j$ belonging to a spiral with parameters $R_{0,j}$, $k_j$, and $\lambda_{0,j}$ the following equalities are valid:
\begin{align}\label{line_p}
 R(\lambda_j) &= |R_{0,j}|e^{k_j(\lambda_j-\lambda_{0,j})},\notag\\
 X_{P_j} &=  R_{0,j} - R(\lambda_j)\cos\lambda_j, \\
 Y_{P_j} &=   R(\lambda_j)\sin\lambda_j \notag
 \end{align}
 (see Eqs. \eqref{R_i} and \eqref{XY_la}). The equation of the straight
line $C_0M$ in Cartesian coordinates is
\begin{equation}\label{line}
Y = (\me R_0 - X)\tan\Lambda.
\end{equation}

Substituting Eq.~\eqref{line_p}  for the coordinates~$X_{P_j}$ and $Y_{P_j}$
into~\eqref{line} gives the equality
\begin{equation}
R(\lambda_j)\sin\lambda_j = [\me R_0 - R_{0,j} + R(\lambda_j)\cos\lambda_j]\tan\Lambda,
\end{equation}
which is valid for point $P_j$ as the intersection of the
spiral and the straight line. Taking into account
Eq.~\eqref{line_p} for $R(\lambda_j)$, we finally obtain the equation for
the longitude $\lambda_j$ of the sought-for point ~$P_j$
\begin{equation}\label{intersect}
|R_{0,j}|e^{k_j(\lambda_j-\lambda_{0,j})}\sin\lambda_j - 
\left[\me R_0 - R_{0,j} + |R_{0,j}|e^{k_j(\lambda_j-\lambda_{0,j})}\cos\lambda_j\right]\tan\Lambda = 0.
\end{equation}
The roots of Eq.~\eqref{intersect} were calculated in the segment
$\lambda_j\in[-\pi;\pi)$. From the two formal roots we chose
the root whose sign coincided with the sign of~$\Lambda$. The
quantity~$\lambda_j$ defines point $P_j$ of Eqs.~\eqref{line_p}.

For fixed~$\Lambda$ we found the median of the set of distances $\left\{R_j\right\}_{j=1}^{N_{\text{sol}}}$, where $R_j=|C_0P_j|$, and two quantiles of the~$R_j$ distribution measured from the median
on different sides of it and containing together a fraction of the distribution for the~$1\sigma$ confidence level, i.e.,
$\frac{\approx68.3\%}{2}$ in each quantile. The outer boundaries of the
quantiles were taken as the boundaries of the confidence interval for the model quantity $R(\Lambda)$ defined
by the model spiral for a given~$\Lambda$ (Fig.~\ref{conf_int}b). Solving
the same problem for various values of~$\Lambda$ allows the
boundaries of the confidence region to be found for a
model spiral with any resolution in~$\Lambda$ (Fig.~\ref{conf_int}b, see
also Fig.~\ref{Per_conf_reg}).

\subsection*{A2. Estimating the Natural Root-Mean-Square
Scatter of Masers \\ Across the Spiral Segment}

\begin{figure}
\hspace{-0.1cm}
\centerline{%
\epsfxsize=6cm%
\epsffile{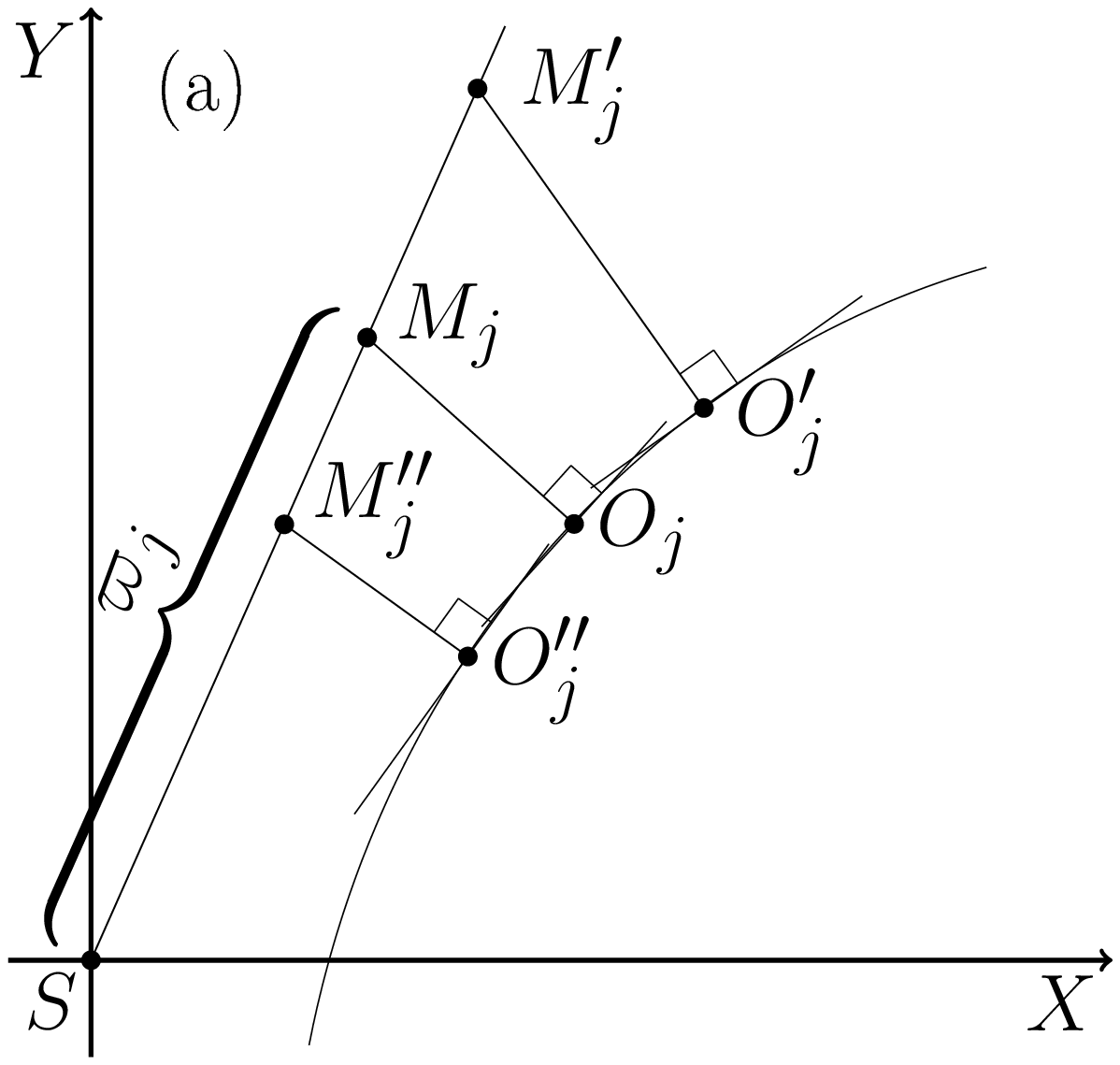}%
\raisebox{-0.11cm}{%
\hspace{1.6cm}%
\epsfxsize=7.1cm\epsffile{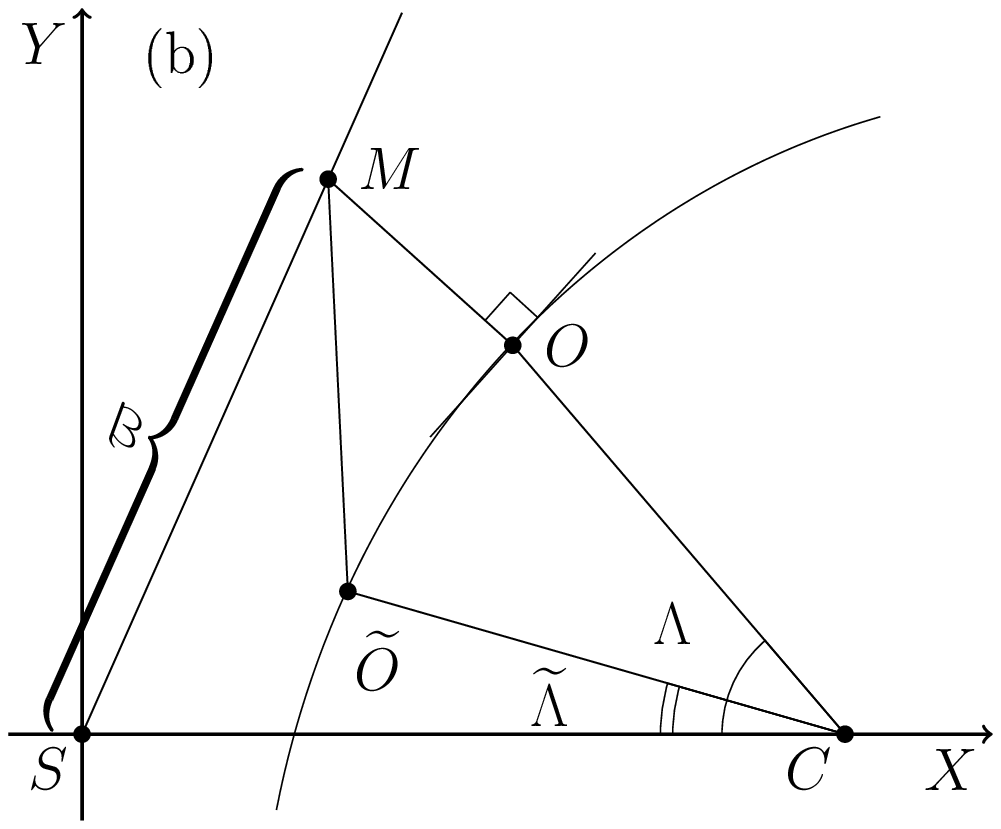}%
}%
}%
\vspace{1em}
\caption{ \rm (a) To the determination of the scatter across the spiral arm segment. (b) To the determination of the distance from a
point $(M)$ to the spiral. $S$ is the position of the Sun, $C$ is the spiral pole.}
\label{width}
\end{figure}

Let $M_j$ be the nominal position of a maser in
accordance with the measurements, $\varpi_j$ and $\sigma_{\varpi_j}$ be,
respectively, the parallax of the maser and its uncertainty given in the catalogue. Here, $j=1, 2, \:\ldots,\:N$,
where $N$ is the size of the sample of masers assigned
to a given segment. Let us introduce the following
notation for the distances from the points on the
line of sight to the feet of the perpendiculars drawn
through these points to the spiral (Fig.~\ref{width}a):
 \begin{align}\label{rho}
   \rho_j &= |M_jO_j|, \quad |M_jS| = \varpi_j^{-1}, \notag \\ 
   \rho_{j,1} &= |M'_jO'_j|, \quad  |M'_jS| = (\varpi_j - \sigma_{\varpi_j})^{-1}, \\
      \rho_{j,2} &= |M''_jO''_j|, \quad  |M''_jS| = (\varpi_j + \sigma_{\varpi_j})^{-1}. \notag
  \end{align}
  Here, $O_j$, $O'_j$, and $O''_j$ are the foot points of the
perpendiculars determined by the positions of points
$M_j$, $M'_j$, and $M''_j$  on the line of sight, respectively; $S$
is the position of the Sun.

Knowing the distances~\eqref{rho}  allows us to obtain
the following estimates for the observed variance of
objects across the spiral arm segment, $\left(\sigma_{\text{w}} \right)_{\text{obs}}^2$, the
contribution of the uncertainties in the parallaxes to
this variance, $\left(\sigma_{\text{w}}\right)_{\varpi}^2$, and the natural variance of objects across the segment, $\left(\sigma_{\text{w}}
\right)_{0}^2$:
\begin{gather}
 \label{sigma_obs}
 \left(\sigma_{\text{w}} \right)^2_{\text{obs}} =  \frac{1}{N-3}\sum\limits_{j=1}^{N} \rho_j^2\,,\\
 \label{sigma_par}
 \left(\sigma_{\text{w}}\right)^2_{\varpi} = 
    \frac{1}{N-3}\left[\sum\limits_{j=1}^{N} \frac{ \left(\rho_j - \rho_{j,1}\right)^2 +   
    \left(\rho_j - \rho_{j,2}\right)^2}{2}\right],\\
 \label{sigma_0}
 \left(\sigma_{\text{w}} \right)^2_{0} =  \left(\sigma_{\text{w}} \right)^2_{\text{obs}} - \left(\sigma_{\text{w}}\right)^2_{\varpi}.
\end{gather}
The distances~\eqref{rho} from the points to the spiral
were calculated as follows. Let $(X_0,Y_0)$ be the Cartesian coordinates of point~$M$;
$\widetilde{O}(\widetilde{X},\widetilde{Y})$ be an arbitrary
point of the spiral with a nominal Galactocentric longitude $\widetilde{\Lambda}$; $R_0$, $k$, and~$\lambda_0$ be the spiral parameters. We found such a value of the longitude $\Lambda$ at which the
distance  $\big|M\widetilde{O}\big|(\widetilde{\Lambda} =
\Lambda)$ was smallest (Fig.~\ref{width}b).

Let us express the coordinates of point $\widetilde{O}$ via~$\widetilde{\Lambda}$ and
the spiral parameters:
\begin{align}\label{tilde_xy}
   \widetilde{X} &= R_0 - |R_0|e^{k(\widetilde{\Lambda}-\lambda_0)}\cos{\widetilde{\Lambda}}, \\
    \widetilde{Y} &=  |R_0|e^{k(\widetilde{\Lambda}-\lambda_0)}\sin{\widetilde{\Lambda}}.
  \end{align}
 Let us find the value of $\Lambda$ that provides the minimum
of the function
 \begin{equation}\
  F(\widetilde{\Lambda}) \equiv \big|M\widetilde{O}\big|^2 = 
    [\widetilde{X}(\widetilde{\Lambda}) - X_0]^2 + [\widetilde{Y}(\widetilde{\Lambda})-Y_0]^2.
 \end{equation}
 After taking the derivative $F'(\widetilde{\Lambda})$, we obtain the equation to determine ${\Lambda}$:
\begin{equation}\label{sp_dist}
   \left(X_0 - R_0 \right)\left(\sin\Lambda - k\cos\Lambda \right) + Y_0\left(k\sin\Lambda + \cos\Lambda \right) - k|R_0|e^{k(\Lambda - \lambda_0)} = 0.
\end{equation}

Having numerically determined the roots $\Lambda$ of
Eq.~\eqref{sp_dist} on the spiral turn $-\pi \leqslant 
\widetilde{\Lambda} < \pi$, we calculate
the corresponding values of $F(\Lambda)$. Since there exist
two extrema of $F(\widetilde{\Lambda})$ on the spiral turn, the smallest of
the two values of~$\sqrt{ F(\Lambda)}$ is the sought-for distance.

\subsection*{A3. The Jackknife Technique}

The jackknife technique is used to estimate the
variance and bias of the estimator (statistical estimate) for a parameter (Quenouille 1949, 1956; Efron and Stein 1981). We mean the bias due to the sample finiteness. Below we give the formulas of the technique as applied to the median estimate $\me R_0$ by the
three-point method as to the estimator.

Let the median $\me R_0$ be found from the set of
triplets formed from a sample of $N$ objects. Consider
N subsamples ($N-1$ in size each) such that the $p$th
object of the original sample is absent in subsample $p$.
For each of the subsamples from all the selected
triplets of its objects we will determine the median
$(\me R_0)_p$\,, $p=1,2,\:\ldots,\:N$. Next, we will calculate the
arithmetic mean of~$(\me R_0)_p$
\begin{equation}
  \langle \me R_0\rangle_J = \frac{1}{N} \sum\limits^{N}_{p=1} (\me R_0)_p\,.
 \end{equation}

The variance of the three-point estimate $\me R_0$ as an
estimator is then
 \begin{equation}
  {\sigma_{R_0, J}^2} = \frac{N-1}{N} \sum\limits^{N}_{p=1} 
    \big[(\me R_0)_p - \langle \me R_0\rangle_J \big]^2,
 \end{equation}
while the corrected value of this estimate is
\begin{equation}
  R_{0,\text{corr}} = N\me R_0 - (N-1)\langle \me R_0\rangle_J\,.
 \end{equation}
 The estimator bias is given by the difference~$\me R_0 - R_{0,\text{corr}}$\,; the correction to the estimate is a bias with the
opposite sign:
\begin{equation}
 \Delta R_0 = R_{0,\text{corr}}-\me R_0\,.
\end{equation}
\end{document}